\documentclass[11pt,english]{article}
\usepackage{graphicx}
\usepackage{amstext}
\usepackage{caption}
\usepackage{etoolbox}
\usepackage{makeidx}
\include{epsf}
\usepackage{amsfonts}
\usepackage{authblk} 
\usepackage{sectsty}
\usepackage{amsmath,amssymb,epsfig}
\usepackage{amscd}
\usepackage{amsthm}
\usepackage{mathrsfs}
\usepackage{dsfont}
\usepackage[applemac]{inputenc}
\usepackage[english]{babel}
\usepackage{enumitem} 
\usepackage[]{latexsym}
\usepackage{braket}
\usepackage{caption}
\usepackage{subfigure}
\usepackage{hyperref}
\usepackage{blindtext}
\usepackage{verbatim}

\hypersetup{
pdftitle={},%
pdfauthor={},%
pdfsubject={},%
pdfkeywords={},%
colorlinks=true,%
linkcolor=blue,%
citecolor=red,%
linktocpage=true,%
hyperfootnotes=true,%
pageanchor=true
}

\newcommand{\partdif}[2]{\frac{\partial #1}{\partial #2}}

\newcommand*{\affaddr}[1]{#1}
\newcommand*{\affmark}[1][*]{\textsuperscript{#1}}

\newtheorem*{proof*}{Proof}

\newcommand{\be}{\begin{equation}}

\newcommand{\ee}{\end{equation}}
\def\beqa{\begin{eqnarray}}
\def\eeqa{\end{eqnarray}}
\def\bean{\begin{eqnarray*}}
\def\eean{\end{eqnarray*}}

\renewenvironment{thebibliography}[1]
         {\section*{References}\frenchspacing\small
          \begin{list}{[\arabic{enumi}]}
         {\usecounter{enumi}\parsep=2pt\topsep 0pt
         \settowidth{\labelwidth}{[#1]}
         \leftmargin=\labelwidth\advance\leftmargin\labelsep
         \rightmargin=0pt\itemsep=1pt\sloppy}}{\end{list}}

 \numberwithin{equation}{section}

\input xy
\xyoption{all}

\usepackage[normalem]{ulem}
\title{\textbf{\textsf{Holographic Signatures of Resolved Cosmological Singularities II: Numerical Investigations}}\vspace{0.35cm}}

\author{
\textsf{Norbert Bodendorfer\affmark[1]\footnote{\texttt{norbert.bodendorfer@physik.uni-regensburg.de}}, Fabio M. Mele\affmark[1]\footnote{\texttt{fabio.mele@physik.uni-regensburg.de}}, and Johannes M\"unch\affmark[1]\footnote{\texttt{johannes.muench@physik.uni-regensburg.de}}}\\
\affaddr{\affmark[1]\textsf{Institute for Theoretical Physics, University of Regensburg,}}\\
\affaddr{\textsf{93040 Regensburg, Germany}}\vspace{-0.5cm}
}

\begin{document}

\maketitle

\begin{abstract}
\textsf{A common strategy to investigate the fate of gravitational singularities in asymptotically AdS spacetimes is to translate the question from the gravitational side to a dual field theory using the gauge/gravity correspondence and to do a field theory computation. Given recent progress in singularity resolution via non-perturbative quantum gravity, it is natural to now turn the question around and to ask about field theory signatures of resolved singularities. An investigation along this line has been initiated in a companion paper, where a finite-distance pole exhibited by the two-point correlator in the dual field theory, which has previously been linked directly to the gravitational bulk singularity, has been resolved in this way. In order to perform analytic computations, some simplifications were necessary. In this paper, we lift these restrictions by tackling the problem numerically. Our analysis shows that the pole in the two-point correlator gets resolved in the same manner as before.}
\end{abstract}

\section{Introduction}

Gravitational singularities are generically occurring in solutions of general relativity \cite{PenroseGravitationalCollapseAnd, HawkingPropertiesOfExpanding}. Their physical interpretation and eventual fate in a theory surpassing classical gravity are puzzling questions that have generated much interest among researchers, see e.g. \cite{NatsuumeTheSingularityProblem, BojowaldSingularitiesAndQuantum} for an overview. It is commonly believed that once a complete quantum theory of gravity is employed, the classical singularities will be resolved. The simplest possible case in which this might happen in a cosmological context is that quantum effects lead to an effective spacetime where the ``big bang'' is replaced by a ``big bounce'', i.e. a quantum regime which interpolates between a contracting and an expanding branch. Such ideas are also of great interest for cosmological applications since they may lead to observable effects, see e.g. \cite{CaiExploringBouncingCosmologies, AshtekarLoopQuantumCosmologyFrom}, and resolve conceptual problems, see e.g. \cite{IjjasBouncingCosmologyMade}.

Within string theory, seen as a theory of quantum gravity, no clear picture of the fate of generic gravitational singularities has emerged so far. In particular, one expects quantum effects to be strong in such a regime, which casts doubt on a straightforward application of perturbative techniques. However, using the AdS/CFT correspondence \cite{MaldacenaTheLargeN, GubserGaugeTheoryCorrelators, WittenAntiDeSitter} as a definition of non-perturbative string theory, the question can be transferred to the dual field theory. Whereas much effort has gone into studying cosmological singularities in this approach, see e.g. \cite{HertogTowardsABig, HertogHolographicDescriptionOf, DasTimeDependentCosmologies, TurokFromBigCrunch, DasCosmologiesWithNull, CrapsQuantumResolutionOf, AwadGaugeTheoryDuals, AwadGaugeTheoriesWith, BarbonAdSCrunchesCFT, SmolkinDualDescriptionOf}, no clear picture has emerged by now.

In order to make progress, it therefore seems reasonable to try to approach the inverse question: given existing results about singularity resolution in models of non-perturbative quantum gravity, is it possible to infer signatures of resolved singularities in the dual field theory? Knowing what to look for, one may be able to confirm these signatures by a field theoretic computation, either analytic or on the lattice. In turn, one could then conclude that non-perturbative quantum gravity can be an adequate description of a subsector of string theory non-perturbatively defined via its dual field theory.

In this paper, we will continue an investigation along this line that was started in a companion paper \cite{BodendorferHolographicSignaturesOf}, which in turn built on the work \cite{EngelhardtHolographicSignaturesOf,EngelhardtFurtherHolographicInvestigations} in the context of the classical gravity approximation in AdS/CFT. In particular, we will lift two key simplifications that were made in \cite{BodendorferHolographicSignaturesOf} in order to be able to compute analytically. They amount to a proper setting of the bulk Planck scale and an asymmetric bounce that has been observed in the non-perturbative context \cite{GuptQuantumGravitationalKasner}. Since both of them affect sensitively the qualitative behaviour in the bulk, their effect on the dual field theory might be a priori non-trivial. Our computations however show that the results of \cite{BodendorferHolographicSignaturesOf} are robust and qualitatively insensitive to these modifications.

\noindent
The paper is organised as follows.\\
In Sec. \ref{setup} we recall the basic setup of \cite{BodendorferHolographicSignaturesOf, EngelhardtHolographicSignaturesOf, EngelhardtFurtherHolographicInvestigations} and their main results. We then outline our numerical strategy to go beyond the results of \cite{BodendorferHolographicSignaturesOf} in Sec. \ref{numericmeth}, which comprises the core of the paper. Our main results are presented in Sec. \ref{sec:Results}. Sec. \ref{conclusion} provides a brief conclusion.

\section{Setup and Previous Results}\label{setup}

\subsection{Classical Setting}\label{setup1}

In a recent series of papers \cite{EngelhardtHolographicSignaturesOf,EngelhardtFurtherHolographicInvestigations}, the AdS/CFT correspondence was used to study holographic signatures of cosmological bulk singularities in the classical gravity approximation. The setup of \cite{EngelhardtHolographicSignaturesOf,EngelhardtFurtherHolographicInvestigations} is the following.  

We consider Kasner-AdS bulk spacetime geometries described by the metric
\begin{equation}\label{clmetric}
ds_5^2=\frac{1}{z^2}\left(dz^2+ds_4^2(t)\right)\;,\qquad\qquad ds_4^2(t)=-dt^2+\sum_{i=1}^3 t^{2p_i}dx_i^2
\end{equation}
\noindent
where we have set the AdS radius to 1. As long as the exponents $p_i$ satisfy the vacuum Kasner conditions $\sum_ip_i=1=\sum_ip_i^2$, Kasner metric $ds_4^2$ is a solution of the 4d vacuum Einstein equations without cosmological constant, while the full metric $ds_5^2$ is a solution of 5d vacuum Einstein equations with negative cosmological constant. It has a curvature singularity at $t=0$. In addition to the translational symmetries in the $x^1, x^2, x^3$ directions, the metric (\ref{clmetric}) is also invariant under the scaling transformation

\begin{equation}\label{scaling}
z\longmapsto\Lambda z,\quad\quad t\longmapsto\Lambda t,\quad\quad x_i\longmapsto\Lambda^{1-p_i}x_i\;.
\end{equation}

Following the AdS/CFT dictionary, the dual description of this bulk system involves $\mathcal N=4$ Super Yang-Mills theory on a Kasner background. Alternatively, by picking a different conformal factor, the metric \eqref{clmetric} can be rewritten as
\be\label{clmetric2}
ds_5^2 = \frac{t^2}{z^2} \left( e^{2\tau} dz^2 -d\tau^2 + \sum_{i=1}^3 e^{-2(1-p_i) \tau} dx_i^2 \right)\;, \qquad t = e^{-\tau}
\ee

\noindent
such that the boundary metric describes an anisotropic deformation of de Sitter space with \eqref{scaling} acting as an isometry on the boundary metric and leaving the conformal factor invariant.

\begin{figure}[t!]
	\centering\includegraphics[width=3.5cm]{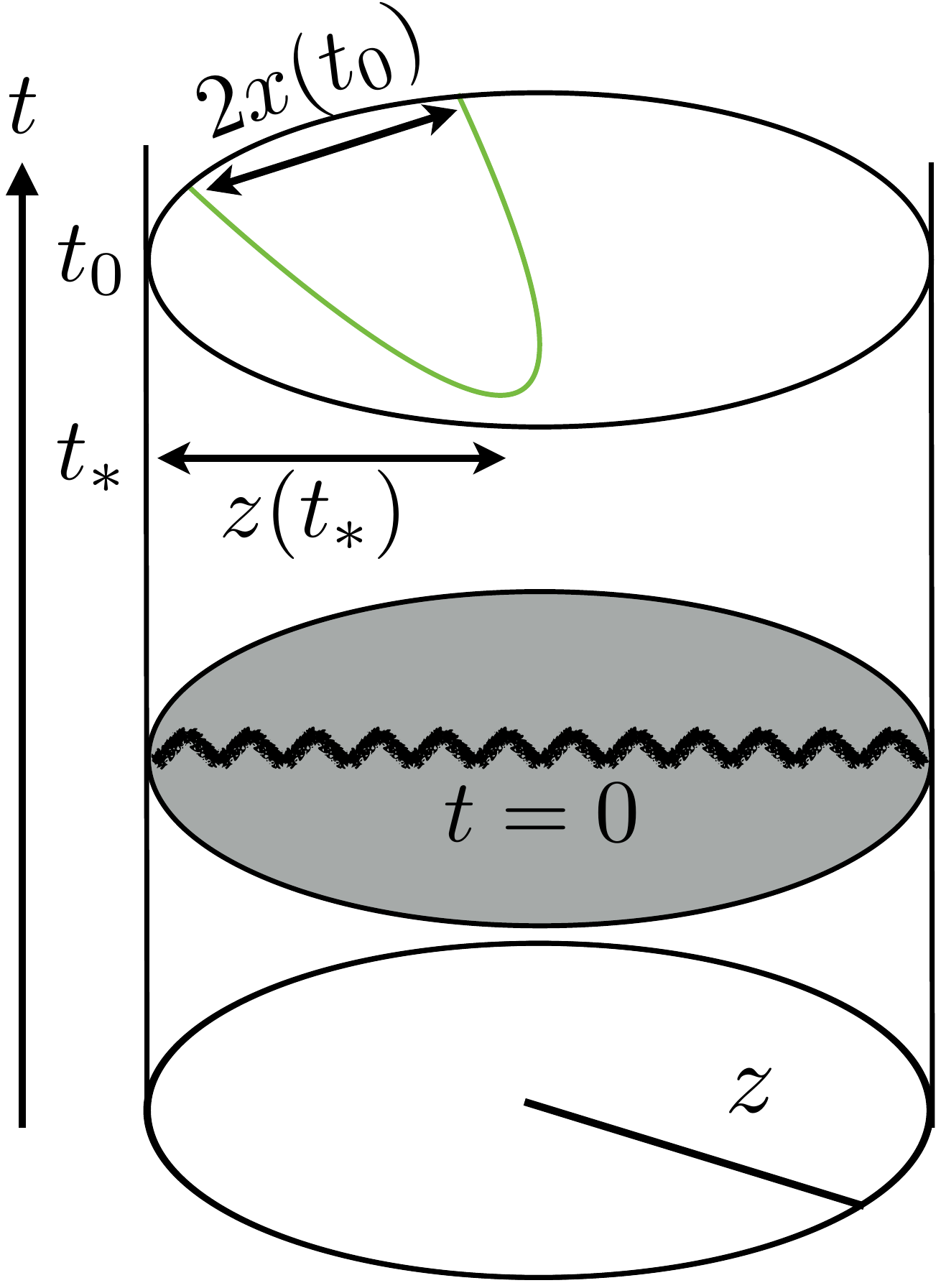}
	\caption{Basic setup to probe bulk singularities by studying the equal time correlator in the dual field theory: in the geodesics approximation the two-point function of a heavy ($m\gg1$) scalar operator $\mathcal O$ is specified by the length of spacelike bulk geodesics anchored at two points on some boundary time slice $t=t_0$.}
	\label{geo}
\end{figure}

In the large $N$ semiclassical bulk limit, the leading contribution to the equal time two-point correlator of a high conformal dimension (heavy) scalar operator $\mathcal O$ is determined by the length of spacelike bulk geodesics connecting two points on some boundary time slice at $t=t_0$. Indeed, in the so-called geodesic approximation\footnote{This is a consequence of the semiclassical WKB approximation according to which the path integral localises to its saddlepoint.} \cite{BalasubramanianHolographicParticleDetection}, the two-point correlator of two heavy scalar operators is dominated by
\be
\label{corr}
\braket{\mathcal O(x)\mathcal O(-x)} \sim \exp{(-\Delta\,L_{\text{ren}})}\;,
\ee
where $\Delta$ is the conformal dimension of $\mathcal O$, which for a $d$-dimensional boundary spacetime is related to the mass $m$ of the bulk field corresponding to the boundary operator $\mathcal O$ via
\be
\Delta=\frac{d}{2}+\sqrt{\frac{d^2}{4}+m^2}\underset{m\gg1}{\simeq}m\;,
\ee
and $L_{\text{ren}}$ is the renormalised (see below) length of a spacelike geodesic connecting the boundary points $(t_0,-x)$ and $(t_0,x)$. Notice that, as sketched in Fig. \ref{geo}, we consider geodesics anchored on the same boundary time slice whose endpoints are separated in only one spatial direction, say $x_1$, hereafter denoted simply by $x$ (and we shall henceforth refer to the corresponding Kasner exponent $p_1$ as $p$). This is due to the fact that geodesics can be thought of as traveling in a (2+1)-dimensional effective spacetime with coordinates $(t,x,z)$ due to the translation symmetry in the $x^i$-directions.

In case of multiple geodesics satisfying given boundary data, a sum over the individual contributions must be included in evaluating the two-point correlator (\ref{corr}). Complex solutions have also to be taken into account as they contribute to the long distance fall-off behaviour $\sim(L_{\text{bdy}})^{-\frac{2\Delta}{1-p}}$ of the two-point correlator for geodesics with \textit{proper} boundary separation $L_{\text{bdy}}$ and not crossing the singularity. Geodesics that propagate in a direction with positive Kasner exponent ($p>0$) are curved away from the singularity, while if we consider the $x$-separation in a negative $p$ direction, geodesics are bent towards the singularity and they are thus characterised by a turning time $t_*<t_0$ for real solutions (see Eq. (3.4) in \cite{EngelhardtFurtherHolographicInvestigations}) and $t_0>0$. In the limit $t_*\rightarrow0$, spacelike bulk geodesics approach a null boundary geodesic for $p<0$ and their tip approaches the bulk singularity. Correspondingly the two-point correlator exhibits a pole at the cosmological horizon scale which is interpreted as a dual signature of the classical bulk singularity. The presence of such a pole indicates that the state in the dual field theory description of the Kasner-AdS metric is not normalisable \cite{EngelhardtFurtherHolographicInvestigations}.

\subsection{Improved Correlator from Effective Bulk Quantum Geometry}\label{setup2}

The possibility that quantum gravity effects might smoothen out this pole and render the two-point correlator finite at non-vanishing spatial separation was already discussed in \cite{EngelhardtFurtherHolographicInvestigations}. However, no explicit mechanism for this was given. Motivated by the discussion in \cite{EngelhardtFurtherHolographicInvestigations}, an example of an improved CFT correlator from quantum gravity effects was provided in \cite{BodendorferHolographicSignaturesOf}. The strategy was to consider effective spacetimes emerging from loop quantum gravity and to repeat the computation of \cite{EngelhardtFurtherHolographicInvestigations} in this context. Specifically, since the metric (\ref{clmetric}) is singular only in its 4d part and the 5d Einstein equations with negative cosmological constant imply that $ds_4^2$ is Ricci-flat, the idea was to keep the $z$-direction classical and to consider a 1-parameter family of quantum corrected metrics for the 4d part, labelled by a parameter $\lambda$ controlling the onset of quantum effects:
\begin{equation}\label{qmetric1}
ds_5^2=\frac{1}{z^2}\left(dz^2+ds_4^2(t)\right),\quad\quad ds_4^2(t)=-dt^2+\frac{a_{ext}^2}{\lambda^{2p}}\left(t^2+\lambda^2\right)^pdx^2+\dots
\end{equation}

\noindent
where dots refer to the other spatial directions which may have different Kasner exponents, $a_{ext}$ is the extremal value of the scale factor, i.e., the value at the bounce ($t=0$). Indeed, for $\lambda>0$ the classical singularity is resolved, while the classical Kasner solution with $a(t)=t^p$ is recovered in the double scaling limit $\lambda\rightarrow0$ and $a_{ext}/\lambda^p\rightarrow1$. We will discuss two important shortcomings of this metric as opposed to a proper motivation from the results of \cite{GuptQuantumGravitationalKasner} below. 

For such kind of bulk metric, the geodesic equations can be solved completely in the $t$-parametrisation. However, the affine parametrisation turns out to be more convenient for computing the renormalised geodesic length. The explicit solution $z(s)$ parametrised with respect to the geodesic length $s$ reads as \cite{BodendorferHolographicSignaturesOf}
\be
\label{zaff1}
z(s)=\frac{z(t_*)}{\cosh{(s-s_0)}}\;,
\ee

\noindent
where we can set $s_0=0$ to start counting the proper distance from the turning point of the geodesic. Eq. (\ref{zaff1}) shows that the length of the geodesics diverges. Therefore, by truncating the geodesics at some boundary regulator $z=\epsilon$, which corresponds to implementing a UV cutoff at energy scale $1/\epsilon$ in the dual field theory, we have
\be
\pm s(z=\epsilon) \stackrel{\epsilon\rightarrow0}{=}\log(2z(t_*))-\log(\epsilon)\;,
\ee

\noindent
from which, subtracting the divergent contribution $-\log(\epsilon)$ originating from geodesics in pure AdS, we get the renormalised geodesic length
\be
\label{Lren1}
L_{\text{ren}}=2\log(2z(t_*)).
\ee
Due to (cfr. Eq. \eqref{corr})
\be
\label{corr2}
\braket{\mathcal O(x)\mathcal O(-x)} \sim (2z(t_*))^{-2\Delta},
\ee
$z(t_*) = 0$ at finite boundary separation corresponds to a finite-distance pole in the two-point correlator.

\begin{figure}[t!]
 \centering
 \subfigure[]
   {\includegraphics[width=7.75cm,height=5.5cm]{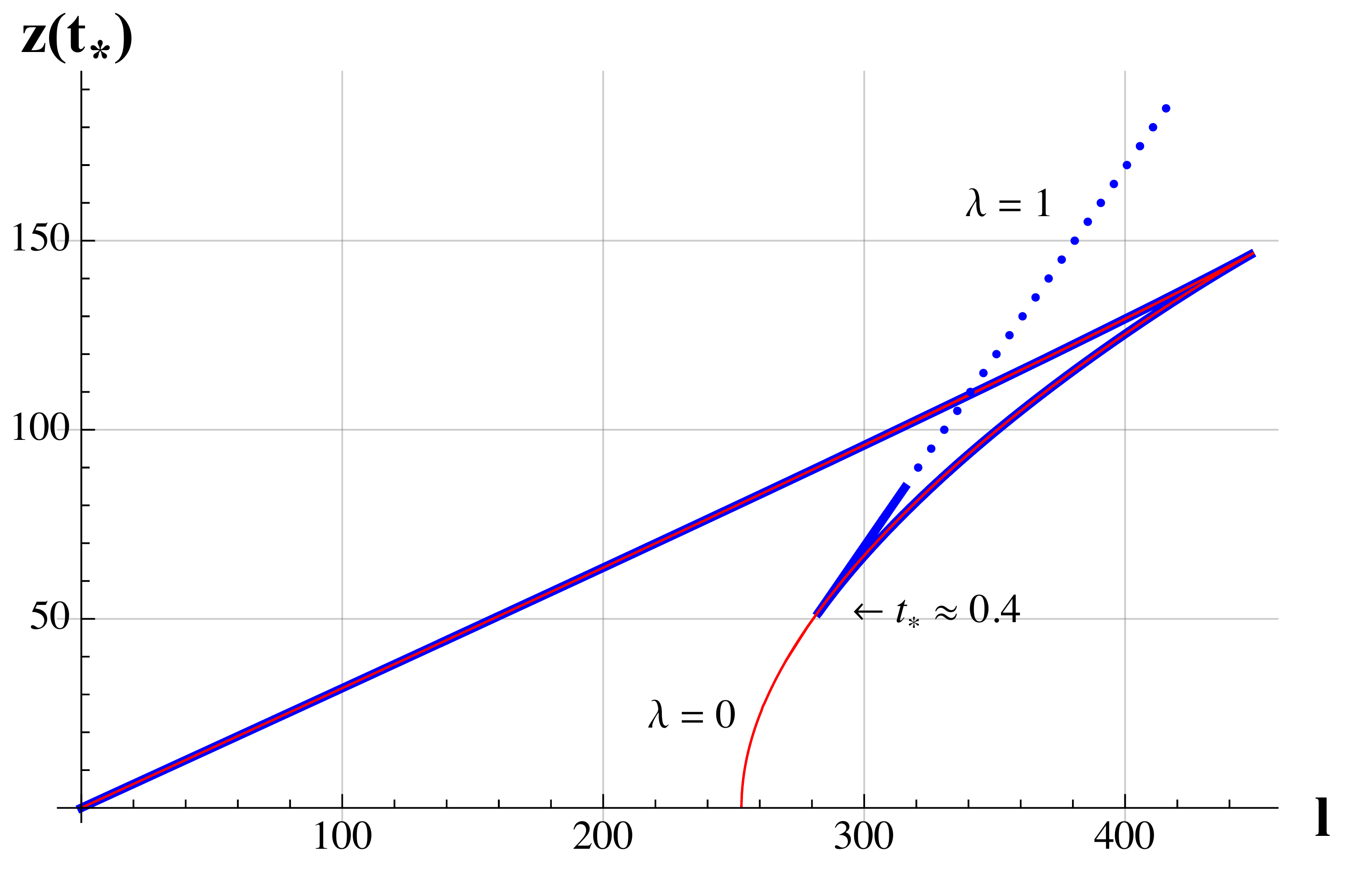}}
 \hspace{2mm}
 \subfigure[]
   {\includegraphics[width=7.75cm,height=5.5cm]{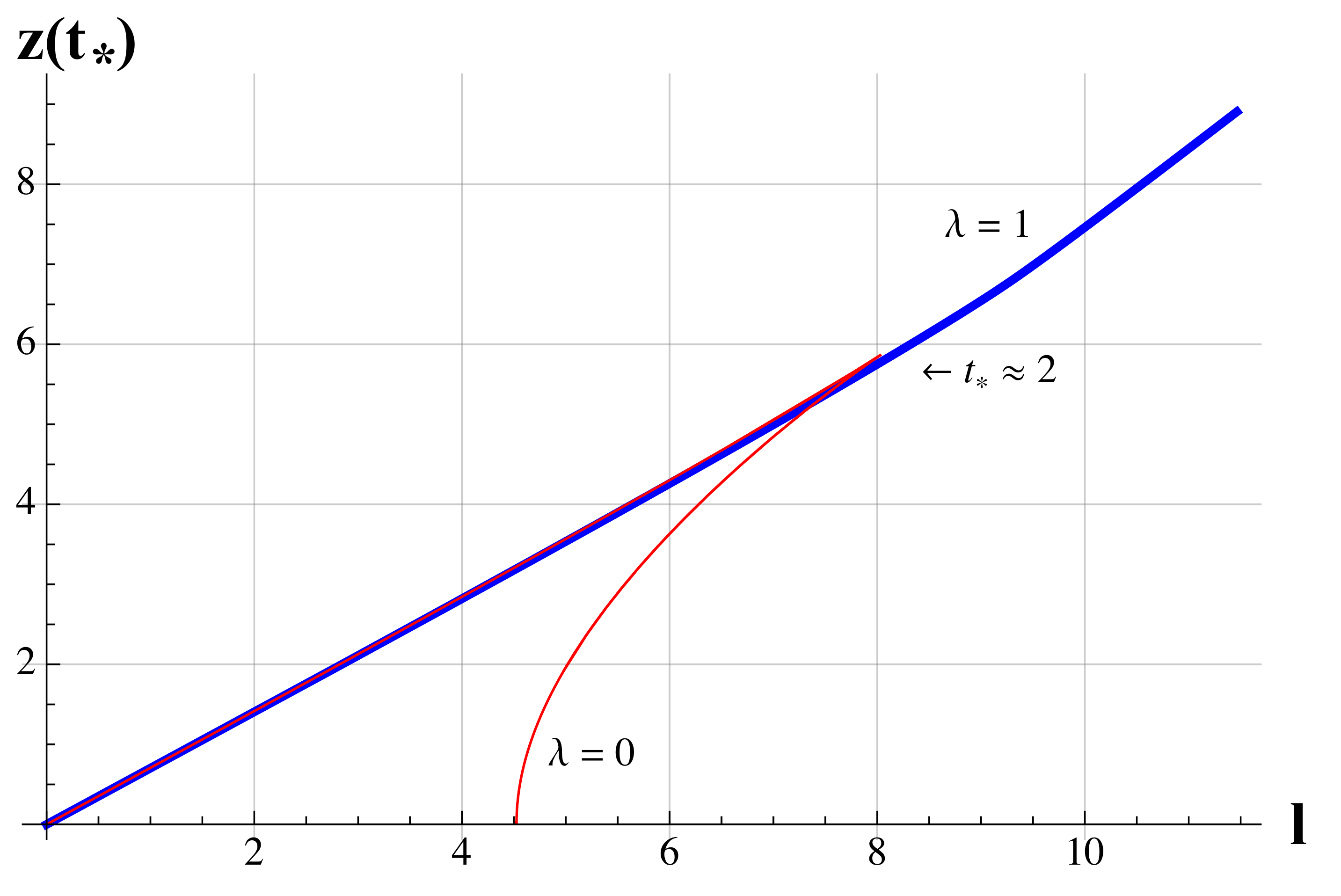}}
 \caption{Plots of $z(t_*)$ vs. $l=x(t_0)$ for $p=-1/4$ and $a_{ext}=1$ taken from \cite{BodendorferHolographicSignaturesOf}. The red line represents the classical case with $\lambda=0$, while the blue line represents the quantum corrected case with $\lambda=1$ respectively for $t_*=t_0=100$ at (0,0) (a), and  $t_*=t_0=4$ at (0,0) (b). The solid blue line was obtained from numerically evaluating the analytic solutions, while the dashed blue line is an asymptotic expansion for $t_*\rightarrow 0$ \cite{BodendorferHolographicSignaturesOf}.}
 \label{an}
 \end{figure}
 
Finally, as shown in Fig. \ref{an}, $z(t_*)$ does not vanish at finite boundary separation for $\lambda=1$ corresponding to the quantum theory (blue curve), unlike the case $\lambda=0$ corresponding to the classical theory (red curve). 
The finite-distance pole in the two-point correlator occurring in the classical theory is thus resolved by quantum effects. There is still a pole at $(0,0)$, but this corresponds to the standard divergence occurring in the coincidence limit. Moreover, there are multiple $z(t_*)$ values corresponding to real geodesics with the same boundary separation (same $x(t_0)$-value), which have to be added in the correlator. The dominant contribution comes from the local minimum of $z(t_*)$ for $\lambda=1$, which represents a clear signature of the resolved classical pole. When $t_0$ gets close enough to $\lambda$, the characteristic behaviour changes (Fig. \ref{an} (b)), but a change of slope in the quantum corrected curve persists and $z(t_*)$ does not vanish at finite distance, i.e., the classical pole is still resolved. Resolution of the finite distance pole along these lines persists for any $\lambda > 0$.

\subsection{Quantum Corrected Metric}\label{5dscale}

The most direct way of investigating holographic signatures of quantum corrected metrics from loop quantum gravity would be to set up 5d quantum Einstein equations based on \cite{BTTI, BTTII, BTTIII} and to extract an effective metric. Unfortunately, this is currently out of technical reach. In order to arrive at a sensible bulk metric, we therefore apply the following key simplification:

As mentioned above, the singularity in the classical bulk metric \eqref{clmetric} originates in the 4d part of the metric. Resolving the singularity in $ds^2_4$ automatically resolves it in $ds^2_5$. The problem can therefore be approximated as one in 4d quantum cosmology, to which much of the effort in loop quantum gravity has gone. In order to ensure that this approximation remains consistent, a possible $z$-dependence of $ds_4^2$ has to be kept small (at the order of $\lambda$) and we will come back to this point.
For now, we remark that the techniques used to construct loop quantum gravity have been employed in a mini-superspace context, leading to the field of loop quantum cosmology \cite{AgulloLoopQuantumCosmology}. Here, a quantum corrected Kasner universe has been studied in \cite{GuptQuantumGravitationalKasner} using effective equations, which were later shown to be in accordance with the results directly obtained from the quantum theory \cite{DienerNumericalSimulationsOf}. Moreover, proposals for how to embed the quantised Bianchi I mini-superspace into full loop quantum gravity have been put forward \cite{AlesciANewPerspective, BodendorferQuantumReductionTo}. The resulting picture is a non-singular bounce that connects two Kasner universes at late times, however with different Kasner exponents before and after the bounce. Following the literature, we will call this feature a Kasner transition.

The onset of quantum effects is again controlled by a parameter $\lambda$ which is related to $\hbar$ and the Barbero-Immirzi parameter \cite{AshtekarQuantumNatureOf}.
A crucial observation for embedding of $ds^2_4$ into $ds^2_5$ as in \eqref{qmetric1} is now that the onset of quantum effects should happen at the 5d Planck scale, and not the 4d scale which $\lambda$ sets. Since the 4d curvature is bounded by $\text{const} / \lambda^2$ in loop quantum cosmology, the relation
\begin{equation}\label{curvature}
R^{(5)}_{\text{Kretschmann}}=R^{(5)}_{\alpha\beta\gamma\delta}R^{(5)\,\alpha\beta\gamma\delta}=z^4R^{(4)}_{\text{Kretschmann}}+\dots\;
\end{equation}
motivates us to substitute $\lambda \mapsto \lambda z$ in order to obtain an onset of quantum effects in the bulk at the 5d Planck scale.

Both of these features, Kasner transitions and a $z$-dependent scale for the onset of quantum effects, were neglected in \eqref{qmetric1} to allow for an analytic computation in \cite{BodendorferHolographicSignaturesOf}. In principle, both of them can have important qualitative effects on the results for the two-point correlator:

First, it was shown in \cite{EngelhardtFurtherHolographicInvestigations} that the divergence in the two-point correlator is due to the bulk geodesic approaching a null geodesic on the boundary. This geodesic is still present in the quantum corrected bulk spacetime since $\lambda z$ goes to zero at the boundary and the metric reduces to classical Kasner. However, our numerical computations suggest that this geodesic is isolated and not the limit of a family of bulk geodesics as in the classical case \cite{EngelhardtFurtherHolographicInvestigations}. 

Second, the long distance behaviour of the correlator in \cite{BodendorferHolographicSignaturesOf} is due to geodesics passing arbitrarily close to $t=0$, where $(t^2+\lambda^2)^p$ in \eqref{qmetric1} has a local maximum (for $p<0$). A Kasner transition as in \cite{GuptQuantumGravitationalKasner} where a negative exponent would transition into a positive one would alter the form of the metric around $t=0$ such that no extremum could be found there.

In the following, we will successively lift these two simplifications and investigate their effect by numerical computations.

\section{Solution Strategy}\label{numericmeth}

As argued in Sec. \ref{setup1} (Eq. \eqref{corr}), the core part in the calculation of the equal time two-point correlator is to compute the geodesic length. For this, we are interested in solving the geodesic equations for a metric of the form
\begin{equation}
ds^2 = \frac{1}{z^2} \left( -dt^2 + a(t,z)^2 dx^2 + dz^2 \right) ,
\end{equation}
\noindent
as a two-point boundary value problem. Indeed, the correlator has to be calculated at a fixed time slice of time $t_0$ with a certain boundary length separation on this time slice given by $L_{\text{bdy}} = 2a(t_0, z=0)x(t_0)$. Because of the translation symmetry in $x$-direction, the coordinate system can always be chosen such that the initial and final point of a geodesic lie symmetrically around the origin. The boundary value problem to solve is then: find all geodesics starting at $(t=t_0, x = -x(t_0), z = 0)$ and ending at $(t = t_0, x = x(t_0), z=0)$, where the input data are $t_0$ and $x(t_0)$\footnote{As already discussed in \cite{BodendorferHolographicSignaturesOf}, the complex solutions to the geodesic equation of the classical metric lie outside of the quantum region and are thus neglected here, as they do not yield new insights. We therefore restrict our computations to real geodesics.}. 
%:

\subsection{Affine Parametrisation and Compactification}

Since the geodesic starts and ends at the same time slice, there should be a turning point in $t$- as well as in $z$-direction. Furthermore, this solution should be symmetric around the turning point, i.e., it has two branches (towards the turning point and back again). A coordinate parametrisation, in particular time parametrisation, would not be a convenient choice  since it does not allow to parametrise both branches at the same time (cfr. \cite{EngelhardtFurtherHolographicInvestigations,BodendorferHolographicSignaturesOf}). This can be achieved by using the affine parametrisation. In such a parametrisation, the geodesic equations is easily derived from the action

\begin{equation}
S = \int ds \sqrt{ g_{\mu \nu}  \frac{dx^{\mu}}{ds} \frac{dx^{\nu}}{ds}} = \int ds \sqrt{-  \frac{\dot{t}^2}{z^2} + \frac{a(t,z)^2}{z^2} \dot{x}^2 + \frac{\dot{z}^2}{z^2} } \;,
\end{equation}
	
\noindent
where dot denotes derivatives with respect to $s$ and we chose the sign appropriate for spacelike geodesics. Using $- \frac{\dot{t}^2}{z^2} + \frac{a(t,z)^2}{z^2} \dot{x}^2 + \frac{\dot{z}^2}{z^2} = 1$, the geodesic equations read

\begin{subequations}\label{eq:geo-s}
	\begin{align}
	&\ddot{t} - \frac{2 \dot{t} \dot{z}}{z} + \frac{\dot{x}^2}{2} \partdif{(a^2)}{t} = 0 \;,
	\\
	\notag
	\\
	&\frac{d}{ds} \left( \frac{a(t,z)^2}{z^2} \dot{x} \right) = 0 \label{eq:I1} \;,
	\\
	\notag
	\\
	&\ddot{z} - \frac{2 \dot{z}^2}{z} + z - \frac{\dot{x}^2}{2} \partdif{(a^2)}{z} = 0 \;.
	\label{geo-eq3}
	\end{align}
\end{subequations}

\noindent
Note that Eq. (\ref{eq:I1}) reflects the constant of motion corresponding to translation symmetry in $x$-direction. Moreover, since the scale factor $a(t,z)$ depends also on $z$, Eq. \eqref{geo-eq3} has an additional term and we cannot explicitly solve for $z$ only as in Eq. \eqref{zaff1}. Therefore, unlike in \cite{EngelhardtFurtherHolographicInvestigations,BodendorferHolographicSignaturesOf}, Eqs. \eqref{eq:geo-s} do not decouple and as such they do not appear to be analytically solvable anymore. 

Thus, we have to reformulate the problem in such a way that it is numerically tractable. Since we want to integrate up to the boundary, which is at infinite proper distance $s$ form the turning point, it is convenient to compactify the parameter $s$ as
\be
\sigma = \tanh(s)  , \quad \sigma \in (-1,1) \;. 
\ee

\noindent
In this parametrisation, the boundary value problem reads

\begin{equation}\label{eq:geo-bvp}
\begin{cases}
t' = p_t & \\  
x' = p_x & \\
z' = p_z & \\
p_t' =  \frac{2 \sigma}{1-\sigma^2} p_t + \frac{2 p_t p_z}{z} - \frac{p_x^2}{2} \partdif{(a^2)}{t}&\\
p_x' = \frac{2 \sigma}{1-\sigma^2} p_x + \frac{2 p_x p_z}{z} - \frac{p_x}{a(t,z)^2} \left( \partdif{(a^2)}{t} p_t + \partdif{(a^2)}{z} p_z \right) &\\
p_z' = \frac{2 \sigma}{1-\sigma^2} p_z + \frac{2 p_z^2}{z} - \frac{z}{(1-\sigma^2)^2} + \frac{p_x^2}{2} \partdif{(a^2)}{z} & \\
&\\
t(-1) = t(1) = t_0 &\\
x(-1) = -x(1) = -l&\\
z(-1) = z(1) = 0 &
\end{cases} \; ,
\end{equation}

\noindent
where prime denotes derivatives with respect to $\sigma$, $l = x(t_0)$, and we introduced $p_t$, $p_x$, $p_z$ to rewrite the equations as first order ODEs. The additional terms in $\sigma$ are due to the reparametrisation properties of derivatives, $\frac{d^2 \sigma}{ds} / \left(\frac{d s}{d \sigma}\right)^2 = -2\sigma/(1-\sigma^2) $ and $\frac{d \sigma}{ds} = 1-\sigma^2$.

\subsection{Mapping Boundary Value Problem into Initial Value Problem}\label{sec:bvp-ivp}

To solve boundary value problems numerically, there are well established methods like the relaxation method (see e.g. \cite{EckerEvolutionOfHolographic,EckerExploringNonlocalObservables}) or the shooting method \cite{PressNumericalRecipes3}. In our case, the problem is more tractable once reformulated as a initial value problem. This makes the numerical solution much simpler and has other advantages, which we will discuss later on. The basic idea is the following: We are only interested in geodesics which have a turning point in $t$ and $z$. Geodesics which do not come back to the boundary or come back but on a different time slice are not solutions of Eqs. (\ref{eq:geo-bvp}). Denoting by $t_*$ and $z_*$ the values of $t$ and $z$ at the turning point, the exact behaviour of the geodesic at this point can be characterised in terms of $t_*$ and $z_*$ only. The turning point itself is then given by the coordinates $(t_*, 0 , z_*)$, where, as already stressed, we can use the translation invariance to fix the value $x_*$ of $x$ at the turning point to be $0$. Furthermore, we can always shift the parametrisation such that the geodesic turns around at $s=0$ $\Leftrightarrow$ $\sigma = \tanh(0) = 0$. At the turning point, the velocities of $t$ and $z$ should vanish, i.e.:
\begin{align}
&0 = \dot{t}(s=0) = t'(\sigma=0) \cdot \frac{d\sigma}{ds}(\sigma=0) = t'(\sigma=0) \;,
\notag
\\
&0 = \dot{z}(s=0) = z'(\sigma=0) \cdot \frac{d\sigma}{ds}(\sigma=0) = z'(\sigma=0) \;.
\notag
\end{align}

\noindent
Using again $- \frac{\dot{t}^2}{z^2} + \frac{a(t,z)^2}{z^2} \dot{x}^2 + \frac{\dot{z}^2}{z^2} = 1$, we find:

\begin{align}
\dot{x}(s = 0) = \frac{z_*}{a(t_*,z_*)} = \frac{z_* \lambda^p}{a_{ext} \left( t_*^2 + \lambda^2 z_*^2 \right)^{\frac{p}{2}}} = x'(\sigma=0) \cdot \frac{d\sigma}{ds}(\sigma=0) = x'(\sigma=0) \;.
\end{align}

\noindent
The boundary value problem \eqref{eq:geo-bvp} can be then rephrased as the following initial value problem
\begin{equation}\label{eq:geo-ivp}
\begin{cases}
t' = p_t & \\
x' = p_x & \\
z' = p_z & \\
p_t' =  \frac{2 \sigma}{1-\sigma^2} p_t + \frac{2 p_t p_z}{z} - \frac{p_x^2}{2} \partdif{(a^2)}{t}&\\
p_x' = \frac{2 \sigma}{1-\sigma^2} p_x + \frac{2 p_x p_z}{z} - \frac{p_x}{a(t,z)^2} \left( \partdif{(a^2)}{t} p_t + \partdif{(a^2)}{z} p_z \right) &\\
p_z' = \frac{2 \sigma}{1-\sigma^2} p_z + \frac{2 p_z^2}{z} - \frac{z}{(1-\sigma^2)^2} + \frac{p_x^2}{2} \partdif{(a^2)}{z} & \\
&\\
t(0) = t_*,\; x(0) = 0,\;  z(0) = z_* &\\
p_t(0) = 0,\; p_x(0) = z_*/a(t_*,z_*), \; p_z(0) = 0  &
\end{cases} \;,
\end{equation}

\noindent
where the initial data $(t_*, 0, z_*)$ and the velocities $(0, z_*/a(t_*,z_*), 0)$ are expressed only in terms of the turning point values $t_*, z_*$, which we give as input. The parameter $\sigma$ runs from $0$ to $1$, that means we only cover one of the two branches. This is not a problem, since around $\sigma = 0$ the solution is symmetric in $t$ and $z$ and anti-symmetric in $x$.

The crucial point is to relate $t_*$ and $z_*$ with the boundary data $t_0$ and $l$. For this, let us define the map
\begin{align}
\quad f\!:\; &\mathbb{R}^2 \longrightarrow \mathbb{R}^2
\notag
\\
\quad\quad &\;(t_*, z_*) \longmapsto f(t_*, z_*) = \left( f^1(t_*, z_*), f^2(t_*,z_*) \right)
\end{align}
\noindent
with
\begin{align}\label{eq:bdymapping}
& f^1(t_*, z_*) = t(\sigma=1) = t_0(t_*,z_*)
\notag
\\
&f^2(t_*, z_*) = x(\sigma=1) = l(t_*,z_*)
\end{align}
\noindent
where $t(\sigma)$, $x(\sigma)$ are solutions of Eq. (\ref{eq:geo-ivp}). 
This map is well-defined due to the uniqueness of solutions to initial value problems. For solving the boundary value problem, it would be necessary to ``invert'' this map\footnote{Proper inversion is not possible, since in general for given $t_0$ and $l$ the boundary value problem can have multiple solutions with different $t_*$ and $z_*$. Hence, $f$ is not injective in general and no inverse map exists.}. Indeed, we are interested in those $(t_*, z_*)$, whose corresponding solutions end at fixed $t_0$. Such points give the level lines of $t_0(t_*,z_*)$ and relate $z_*$ with $t_*$ (see Eq. (5.2) in \cite{EngelhardtFurtherHolographicInvestigations} for such a relation in the classical case). This can be formalised by considering a curve $c_{t_0}^{\mu}(\tau) = (t_*(\tau), z_*(\tau))^{\mu}$, parametrised by $\tau$, which satisfies:

\begin{equation}
\frac{d c_{t_0}^{\mu}}{d\tau} \; \partdif{}{x^{\mu}} t_0(t_*(\tau), z_*(\tau)) = 0, \quad \quad x^{\mu} = (t_*, z_*)^{\mu} \;.
\end{equation}

\noindent
Similarly, the level lines $c_{l}^{\mu}(\tau)$ of $l$ satisfy 

\begin{equation}
\frac{d c_{l}^{\mu}}{d\tau} \; \partdif{}{x^{\mu}} l(t_*(\tau), z_*(\tau)) = 0, \quad \quad x^{\mu} = (t_*, z_*)^{\mu} \;.
\end{equation}

\noindent
The boundary value problem Eq. \eqref{eq:geo-bvp} is then solved by all intersection points $(t_*^i,z_*^i)$, $i = 1,2,\dots$ of the two level lines. 

The boundary value problem can therefore be reformulated in terms of an initial value problem, the calculation of two level lines, and the calculation of their intersections. For doing so, we have to solve the initial value problem for numerous values of $t_*$ and $z_*$. For our problem this is convenient since we are not interested in specific values of $l$. Instead, we want to vary $l$. Furthermore, we are in principle interested in different values of $t_0$. This means it is possible to solve the initial value problem for a given grid of $t_*$ and $z_*$ and use this data to calculate different $t_0$-level lines, which decreases the total numerical effort\footnote{This method can be compared with the shooting method. There, an initial value problem is solved for different values of the velocities, until the second boundary point is hit. The idea is now to not throw away the ``misshots'' and instead remember the corresponding velocities as a solution of a different boundary value problem. Once one is interested in different boundary values, this saves computational effort.}.

Numerically, we implemented this method with use of Matlab and its built in library. For the solution of the ODE we used the routine \texttt{ode45}, which is based on fifth-order Runge-Kutta method with adaptive step size. For the calculation of the level lines we used the routine \texttt{contourc}. Since the coefficients of Eqs. \eqref{eq:geo-ivp} diverge at $\sigma=\pm1$, we cut the integration interval such that $\sigma \in [0, 1-\epsilon]$, where $\epsilon = 0.00001$. 
We have performed extensive cross-checks on the numerics, including the reproduction of the analytical results of \cite{BodendorferHolographicSignaturesOf}.

\subsection{Renormalised Geodesic Length and Two-Point Correlator}\label{sec:Lren}

For calculating the two-point correlator, it is necessary to compute the renormalised geodesic length. We are interested in the length of the complete geodesic, i.e. from boundary to boundary. In affine parametrisation, with $s=0$ at the turning point, this is $L = 2s$. However, since by construction the conformal boundary lies at infinity, this value diverges and a renormalisation procedure is necessary. 
The geodesic length is thus only evaluated up to a small value $z_{UV}/t \rightarrow 0$. This ratio is preserved by the scaling symmetry \eqref{scaling} and, as discussed in \cite{EngelhardtFurtherHolographicInvestigations}, a constant value of the conformal factor in the metric  \eqref{clmetric2} corresponds to a constant UV cutoff in pure AdS. Then the geodesic length of a pure AdS-geodesic is subtracted and the limit, which should be finite, is taken. Numerically, this is not a trivial task, since the analytical expression is not available and hence the singular part cannot be isolated (cfr. Eq. \eqref{Lren1}). Nevertheless, the strategy is the same: Define $L_{\text{ren}} = L - L_0$, where $L$ and $L_0$ are evaluated at a $z$-cutoff $z_{UV}$ and $L_0$ is a geodesic in pure AdS (see also \cite{EckerEvolutionOfHolographic, EckerExploringNonlocalObservables}). The final result should be independent of $z_{UV}$ and hence coincide in the limit $z_{UV}/t \rightarrow 0$. Let us recall that the range of $\sigma$ is $[0,1]$\footnote{In practice we never reach 1, because of the cutoff $\epsilon$ (see Sec. \ref{sec:bvp-ivp}). $\epsilon$ is chosen small enough to not conflict with the following.} and that $s$ and $\sigma$ are related by $s=\text{arctanh}\left(\sigma\right)$. 
Hence, as expected, $s$ diverges for $\sigma \rightarrow 1$. To renormalise the geodesic length, the solution is evaluated up to the given $z_{UV}$ and the corresponding value of the curve parameter $\sigma$ is read off, say $\bar{\sigma}$ such that $z(\bar{\sigma})=z_{UV}$. With this cutoff, we have
\be
\sigma\in[0,1-\delta] \quad, \quad \delta=1-\bar\sigma \ll 1 \;,
\ee

\noindent
where the condition $\delta \ll 1$ is satisfied as long as $z_{UV}/t \ll z_*/t_*$. Subtracting the divergent part of a pure AdS-geodesic, which is $-\log(\vert z_{UV}/t(\bar{\sigma}) \vert)$ for a single branch, this gives the renormalised length
\be\label{renlen}
L_{\text{ren}}=2\text{arctanh}\left(1-\delta\right)+2\log{\left(\left|\frac{z_{UV}}{ t(\bar{\sigma}) }\right|\right)} \;.
\ee

\noindent
To check that the limit $z_{UV}/t \rightarrow 0$ (i.e. $\bar{\sigma} \rightarrow 1$, $\delta \rightarrow 0$) is finite, let us notice that
\begin{equation}
\text{arctanh}\left(1-\delta\right)+\log\left(\left|\frac{z_{UV}}{t(\bar{\sigma})}\right|\right)=-\frac{1}{2}\log{\left(\frac{\delta \cdot t(\bar{\sigma})^{2}}{z_{UV}^{2}} \right)}+\frac{1}{2}\log{\left(2\right)}+\mathcal O(\delta) \;,
\end{equation}

\noindent
and, as can be checked numerically, for small values of $z_{UV}$, the quantity \eqref{renlen} approaches a non-zero constant.

In order to conclude that the two-point correlator is non-singular, we need to check that $L_{\text{ren}}$ remains finite. Let us recall from Sec. \ref{setup} that both in the classical case \cite{EngelhardtHolographicSignaturesOf,EngelhardtFurtherHolographicInvestigations} and the quantum corrected case considered in \cite{BodendorferHolographicSignaturesOf}, the relation was\footnote{Unlike Eq. \eqref{Lren1}, we have a $t_0$ factor in the logarithm. This difference stems from regulating with $z_{UV} \rightarrow 0$ instead of $z_{UV}/t \rightarrow 0$ in \cite{BodendorferHolographicSignaturesOf} where also $t_0$ was settled to $1$.}

\begin{equation}\label{eq:Lrenofzstar}
\frac{L_{{\text{ren}}}}{2} = \log\left(\frac{2z_*}{t_0}\right) \;.
\end{equation}
\noindent
 As discussed in Sec. \ref{setup2}, this corresponds to the $z_*$-dependence \eqref{corr2} in the two-point correlator. Using the method of this section we are able to relate numerically $z_*$ and $l$. 
We expect the same or a similar dependence also in the cases under consideration. What changes is the relation $z_*(l)$, which differs from case to case. The results of the numerical calculation are presented in the next section.

\section{Results} \label{sec:Results}

\subsection{5d Planck Scale}\label{sec:metric1}

We first applied this method to the metric 
\begin{equation}\label{qmetric2}
ds_5^2=\frac{1}{z^2}\left(dz^2-dt^2+\frac{a_{ext}^2}{\lambda^{2p}}\left(t^2+\lambda^2z^2\right)^pdx^2+\dots\right)\;,
\end{equation}
which, as can be checked by calculating the Kretschmann scalar, features an onset of quantum gravity effects at the 5d Planck scale (cfr. Sec. \ref{5dscale}) but neglects Kasner transitions. For $t \gg z \lambda$, where quantum corrections are negligible in \eqref{qmetric2}, the classical Kasner-AdS solution of the 5d-Einstein equations is recovered. The $z$-derivative of $a(t,z)$ is $\mathcal O(\lambda)$ with finite coefficients and can thus be accounted for by quantum corrections in the $z$-direction, which we systematically neglect here. Also, the proper classical boundary limit exists. These points will be relevant and highly non-trivial also in Sec. \ref{kasner-trans} where Kasner transitions are included.

For this metric the derivatives of $a(t,z)^2$ entering Eqs. \eqref{eq:geo-ivp} are given by

\begin{align}
&\partdif{(a^2)}{t} = 2 p \frac{a_{ext}^2}{\lambda^{2p}} t \left( t^2 + \lambda^2 z^2 \right)^{p-1} = 2 a(t,z)^2 \frac{p t}{t^2 + \lambda^2 z^2} \;,
\\
&\partdif{(a^2)}{z} = 2 p \frac{a_{ext}^2}{\lambda^{2p}} \lambda^2 z \left( t^2 + \lambda^2 z^2 \right)^{p-1} = 2 a(t,z)^2 \frac{p \lambda^2 z}{t^2 + \lambda^2 z^2} \;.\label{zderivative}
\end{align}

\begin{figure}[t!]
	\centering
	\subfigure[]
	{\includegraphics[width=7.75cm,height=5.6cm]{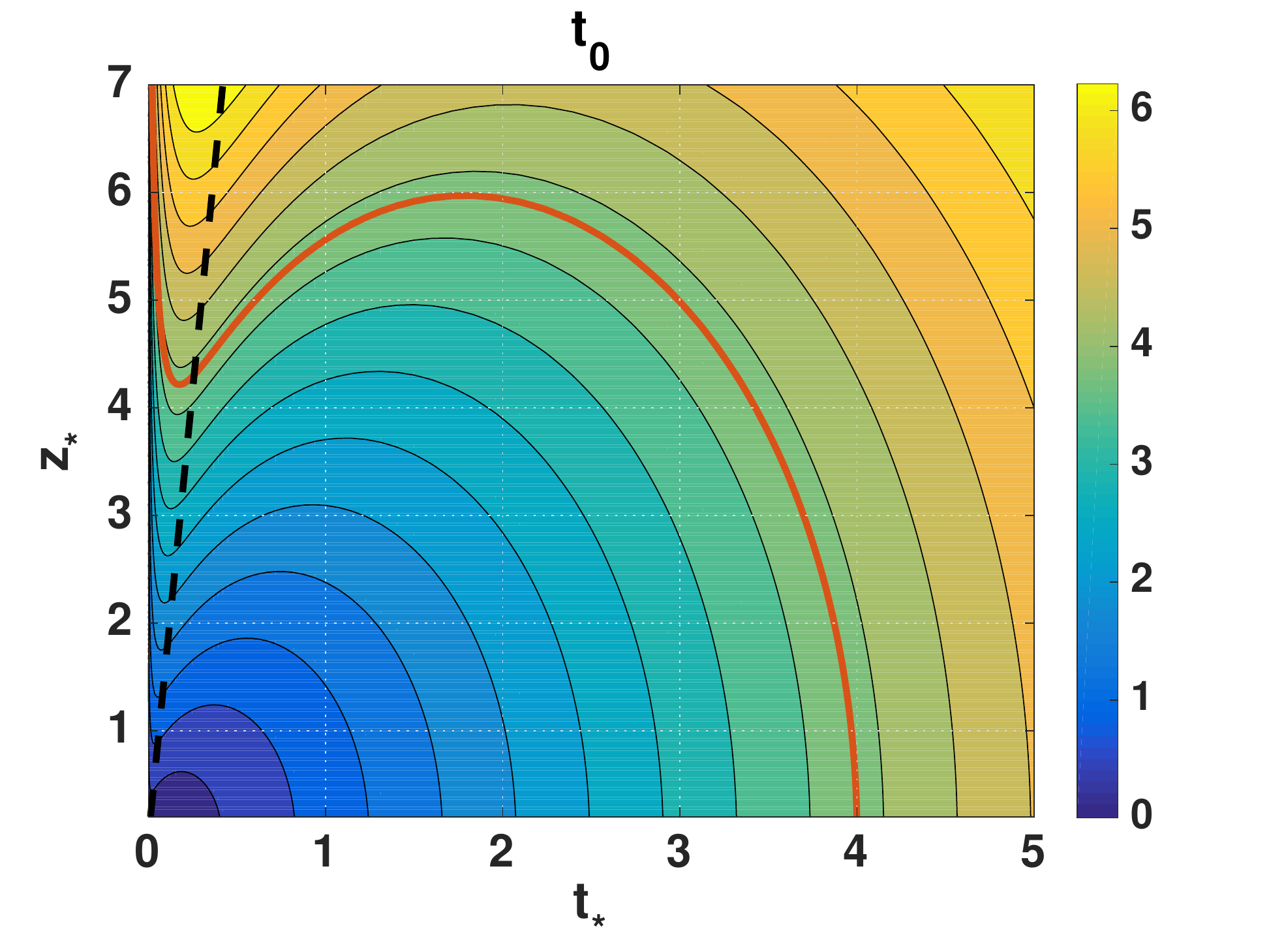}}
	\hspace{2mm}
	\subfigure[]
	{\includegraphics[width=7.75cm,height=5.5cm]{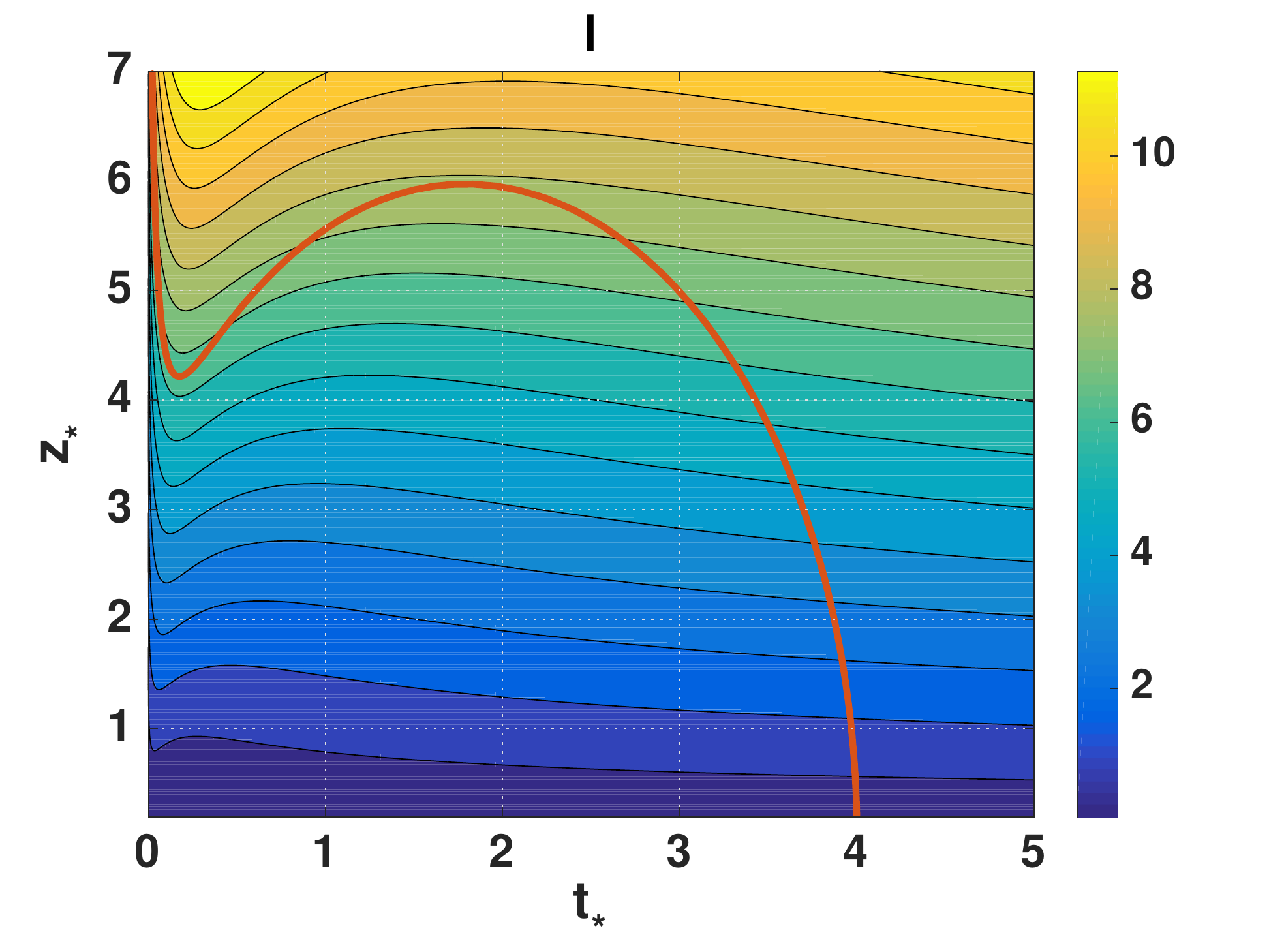}}
	\caption{Colour plot of (a) $t_0(t_*,z_*)$ and (b) $l(t_*,z_*)$ for $p = -1/4$, $\lambda = 0.06$, $a_{ext} = \lambda^p$. The red curve corresponds to the $t_0 = 4$ level line. This contour is also plotted in (b). The black dashed line corresponds to $t^2 = \lambda^2 z^2$ and separates quantum and classical regime.}
	\label{fig:contour1}
\end{figure}

\begin{figure}[t!]
	\centering
	\subfigure[]
	{\includegraphics[width=7.75cm,height=5.6cm]{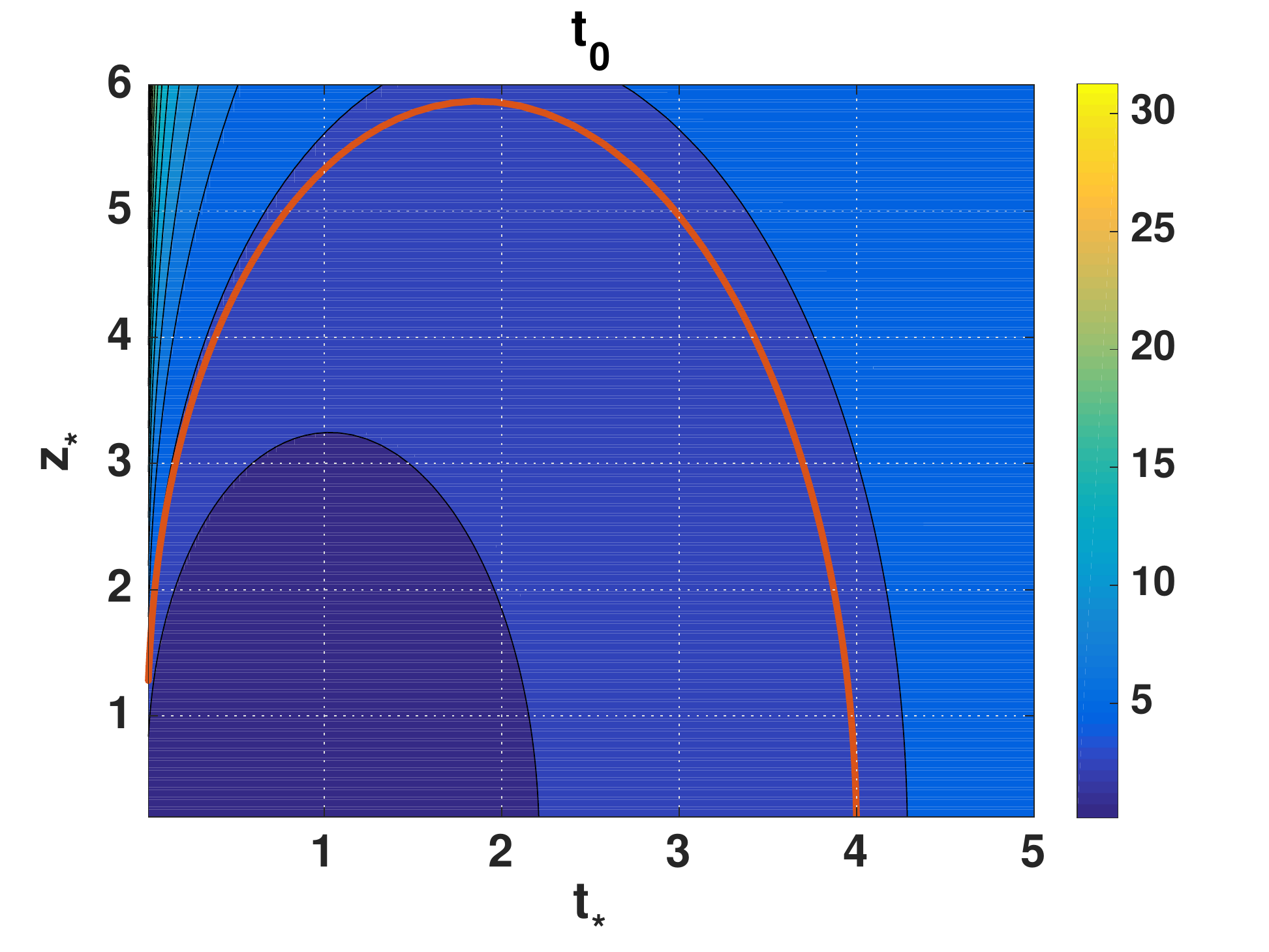}}
	\hspace{2mm}
	\subfigure[]
	{\includegraphics[width=7.75cm,height=5.5cm]{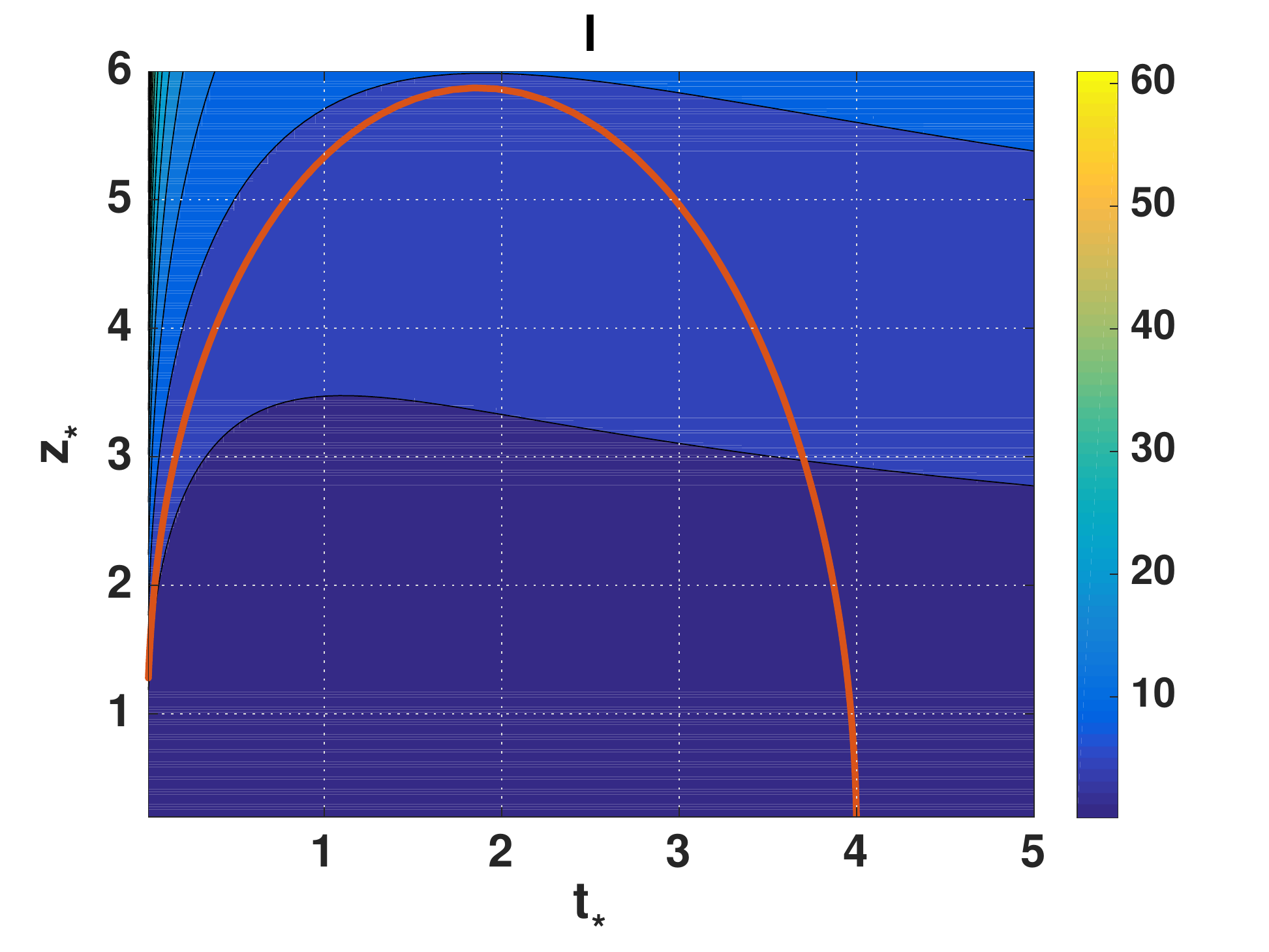}}
	\caption{Colour plot of (a) $t_0(t_*,z_*)$ and (b) $l(t_*,z_*)$ for the classical Kasner-AdS metric ($\lambda = 0$ in \eqref{qmetric2}) with $p = -1/4$. The red curves correspond to the $t_0 = 4$ level line.}
	\label{fig:contour1_class}
\end{figure}

\noindent
The solutions of $t_0(t_*,z_*)$ and $l(t_*,z_*)$ describe surfaces in a 3d space spanned by $(t_*,z_*,t_0)$ and $(t_*, z_*, l)$, respectively. To visualise them we report $z_*$ vs. $t_*$ in Fig. \ref{fig:contour1} where the third direction (respectively $t_0$ and $l$) is replaced by a colour scale.
For this calculation we fixed the parameters to $p = -1/4$ and $\lambda = 0.06$. The range of $t_*$ and $z_*$ is chosen to be between $[0,10]$. Let us focus on Fig. \ref{fig:contour1} (a) first. Among the level lines corresponding to different values of $t_0$, we selected for instance the one for $t_0=4$ (red curve). This level line relates $z_*$ and $t_*$ for that given constant value of $t_0$. We can compare this now with the classical case (see Fig. \ref{fig:contour1_class}).
The classical region is in the area where $t^2 \gg \lambda^2 z^2$, but since $\lambda = 0.06$ is chosen very small the ``dividing line'' $t^2 = \lambda^2 z^2$ is close to the $z_*$-axis (see black dashed line in Fig. \ref{fig:contour1} (a)). Indeed for large $t_*$ (here $\gtrsim 1$) and small $z_*$ (here $<10$) we see exactly the classical behaviour (cfr. Fig. \ref{fig:contour1_class}). On the other hand, going to the quantum regime ($t^2\ll \lambda^2 z^2$), i.e. close to the $z_*$-axis, we see that the level lines exhibit turning points. Therefore, unlike the classical case where the finite-distance pole in the two-point correlator was due to bulk geodesics approaching a null geodesic lying entirely on the boundary ($z_* \rightarrow 0$ for $t_* \rightarrow 0$ on a constant $t_0$ level line), quantum corrections of the metric induce a turning point which leads to a growing $z_*$ for $t_* \rightarrow 0$, thus showing, within our numerical accuracy ($t_*\gtrsim10^{-8}$), that this null-boundary solution is isolated and not the limit of a family of bulk geodesics.

\begin{figure} [t!]
	\centering
	\includegraphics[width=9cm]{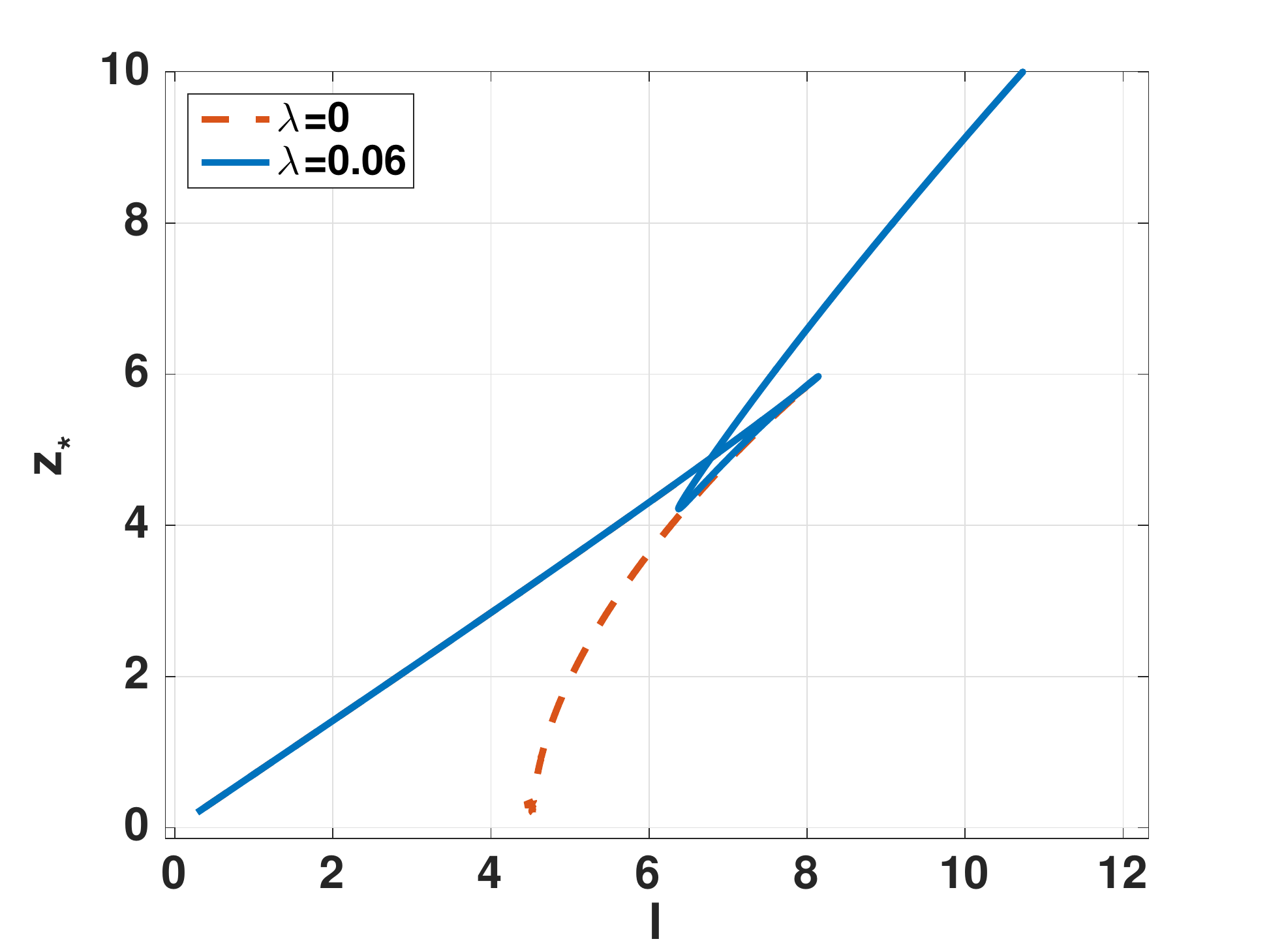}
	\caption{Plot of $z_*$ vs. $l$ for $t_0=4$ and $p = -1/4$, $\lambda = 0.06$, $a_{ext} = \lambda^p$. The blue curve corresponds to the quantum corrected metric \eqref{qmetric2}, the red curve to the classical metric ($\lambda = 0$ in \eqref{qmetric2}).}
	
	\label{fig:zstarvsl1}
\end{figure}

The contour plot of $l$ is reported in Fig. \ref{fig:contour1} (b) where we also included the $t_0 = 4$ level line shown in Fig. \ref{fig:contour1} (a) (red curve). As we see in the plot, $l$ increases as $z_*$ becomes larger. Following the level line and reading off the corresponding values of $l$ leads to the plot of Fig. \ref{fig:zstarvsl1}, where we plotted both the quantum corrected metric (blue line) and the classical result (red line).
The classical result agrees with the results of \cite{EngelhardtHolographicSignaturesOf,EngelhardtFurtherHolographicInvestigations} (see red line in Fig. \ref{an}). According to the analytical results, the red line should hit $z_*=0$ for finite $l$. However, since we introduced a cutoff in $z_*$, the classical red line does not reach the $z_*=0$ axis. Close to the cutoff, some numerical uncertainties occur and we are not able to see the classical pole. Nevertheless, the classical curve converges towards $z_*=0$ for finite $l$ within numerical accuracy and hence, as argued in Sec. \ref{setup}, this leads to a pole in the two-point correlator of the boundary theory. Concerning the quantum case (blue line), we see a turning point in $z_*(l)$ at finite $l$, and then $z_*$ increases again as it was already visible in Fig. \ref{fig:contour1}. Therefore, $z_*$ never hits $0$ for finite non-zero values of $l$. Moreover, as in the analytical case (Fig. \ref{an}), there are multiple solutions corresponding to the same boundary separation, whose contribution has to be added in the two-point correlator. Note that because of the cutoff in $z_*$ both the classical and quantum curves do not start at $z_*=l=0$.

\begin{figure} [t!]
	\centering
	\includegraphics[width=9.75cm]{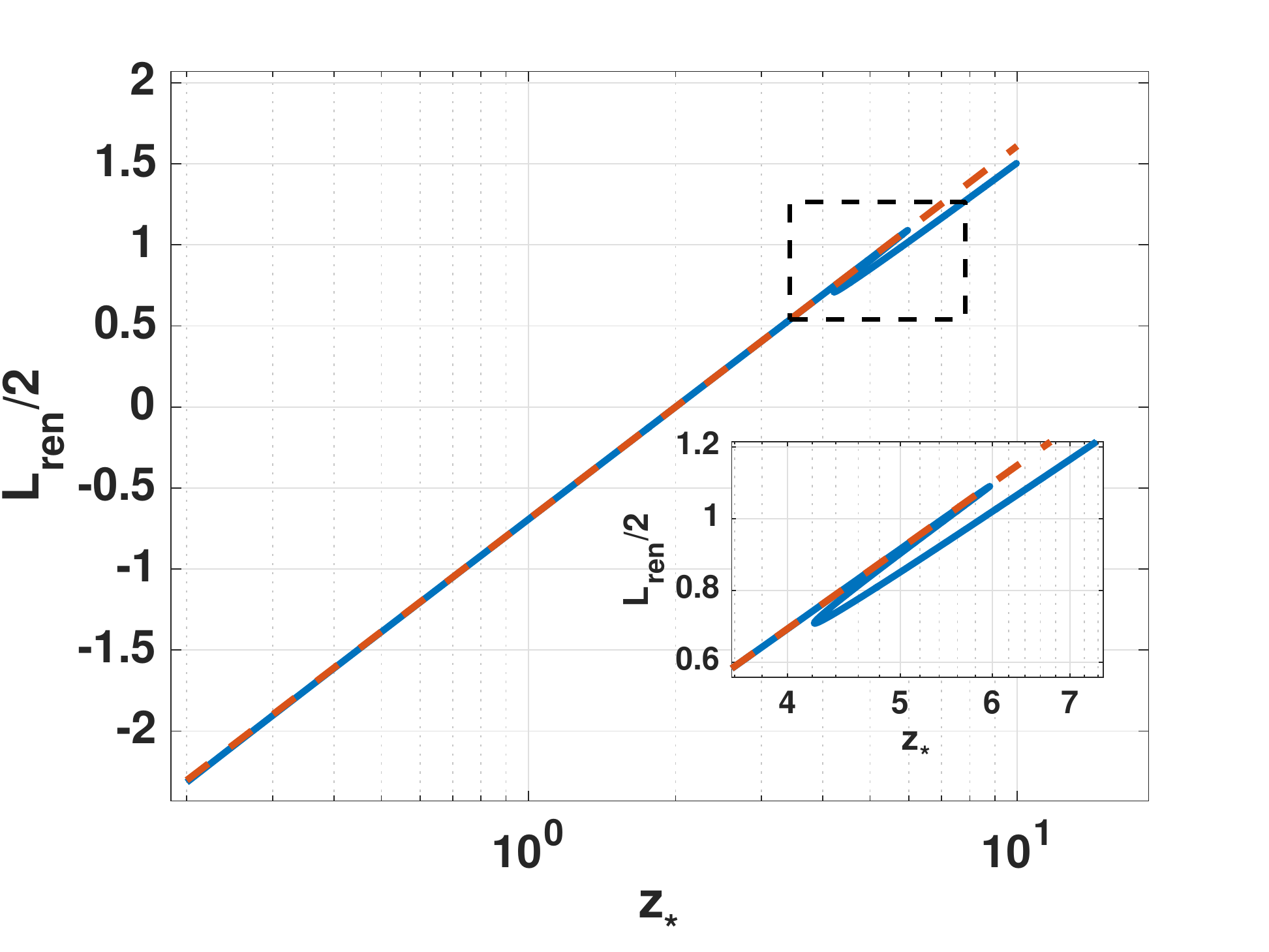}
	\caption{Plot of $L_{\text{ren}}$ on y-axis vs. $z_*$ on a logarithmic x-axis for the metric \eqref{qmetric2} (blue) with $t_0 = 4$, $\lambda = 0.06$, $p=-1/4$, $a_{ext} = \lambda^p$. The almost linear behaviour indicates a $\log$-dependence of $L_{\text{ren}}$ from $z_*$ like Eq. \eqref{eq:Lrenofzstar} (red dashed).}
	
	\label{fig:Lrenvszstar1}
\end{figure}

The next step is to calculate $L_{\text{ren}}$ by means of the procedure described in Sec. \ref{sec:Lren}. A possible dependence like Eq. \eqref{eq:Lrenofzstar} can be easily visualised in a log-plot, where a straight line is expected (red dashed line in Fig. \ref{fig:Lrenvszstar1}). As shown in Fig. \ref{fig:Lrenvszstar1}, our numerical solutions (blue line) exhibit such a dependence.
Nevertheless, there are some subtleties to be discussed. First of all, the upper region provides a purely quantum contribution to the long distance behaviour of the two-point correlator. As can be checked, in agreement with the results of \cite{BodendorferHolographicSignaturesOf}, this contribution decays faster than the lower region (short distance) contribution. Moreover, there are turning points in the line which reflects the above mentioned existence of multiple solutions for a given boundary separation. Indeed, the renormalised length $L_{\text{ren}}$ is calculated for each point along the same $t_0 = 4$ level line selected before in Fig. \ref{fig:contour1}. The turning points are also present in Fig. \ref{fig:Lrenvszstar1}, where some values of $z_*$ are passed more than once. Nevertheless, the shape of a straight line is kept. In the lower region, which contributes to the short distance behaviour of the two-point correlator, deviations from this log-dependence occur below our chosen range of evaluation, but these are just numerical artifacts. Indeed, in this region, $z_*$ comes close to $z_{UV}$, the approximation $\frac{z_{UV}}{t}\ll \frac{z_*}{t_*}$ fails, and the error increases. However, our main interest concerns the behaviour $L_{\text{ren}} \propto \log(z_*)$, or in other words, that $L_{\text{ren}}(z_*)$ is well-behaved in the sense that it does not diverge at finite values of $z_*$. Together with the above results, we can finally conclude that the resolution of the singularity in the bulk also resolves the singularity of the two-point correlator of the boundary theory when the onset of quantum gravity effects happens at the 5d bulk Planck scale.

\subsection{Inclusion of Kasner Transitions}\label{kasner-trans}

We did not find a completely satisfactory 5d metric that incorporates 4d Kasner transitions as found in loop quantum cosmology, the reason for which will be explained below. For the purpose of this paper, which is to show that the finite-distance pole in the two-point correlator can be resolved by quantum gravity effects, we will ignore this point and simply conclude that for two possible straightforward proposals for a metric incorporating Kasner transitions, the pole remains resolved. While this does not settle the issue in that we do not have access to a completely satisfactory 5d effective metric, it again supports and strengthens our previous results.

Let us first consider the following 5d quantum corrected bulk metric

\begin{equation}\label{qmetric3a}
ds_5^2=\frac{1}{z^2}\left(dz^2-dt^2+a(t,z)^2dx^2+\dots\right)
\end{equation}
with

\begin{equation}\label{qmetric3b}
a(t,z)^2=\frac{a_{ext}^2}{\lambda^{2p}}\left(t^2+\lambda^2 z^2\right)^p\exp{\left[2\Delta p\sinh^{-1}\left(\frac{t}{z\lambda}\right)\right]},\qquad\Delta p\in\mathbb R\;.
\end{equation}

The explicit form of the metric (\ref{qmetric3b}) has not been derived from any specific quantum gravity model. Nevertheless, our choice can be motivated as follows. We first consider a 4d Planck scale quantum corrected bulk metric with
\be
\label{CM}
a(t)=\frac{a_{ext}}{\lambda^{p}}\left(t^2+\lambda^2\right)^{p/2}\exp{\left[\Delta p\sinh^{-1}\left(\frac{t}{\lambda}\right)\right]}, 
\ee
which implements a smooth transition between two Kasner universes at late and early times (more details below).

The four-dimensional part of this metric can be directly related to the metric proposed by Chamseddine and Mukhanov \cite{ChamseddineResolvingCosmologicalSingularities} in their modified version of General Relativity that implements the idea of a limiting curvature $\epsilon_m$ if we identify $p=1/3$, $\lambda=1/\sqrt{3\varepsilon_m}$, and specify the values of $a_{ext}$ and $\Delta p$ accordingly (the explicit expressions are not relevant for our present purposes but they can be easily derived by direct comparison with Eq. (40) in \cite{ChamseddineResolvingCosmologicalSingularities}). Here however, we do not specify the values of $p, a_{ext}$ and $\Delta p$ leaving them as generic input parameters in our numerical analysis. The Chamseddine-Mukhanov model \cite{ChamseddineResolvingCosmologicalSingularities} has been proposed as a toy model for an effective theory of quantum gravity in \cite{BodendorferOnTheCanonical, LangloisEffectiveLoopQuantum} by showing that it agrees with the effective dynamics of loop quantum cosmology (LQC) in the spatially flat, homogeneous and isotropic sector if one identifies the limiting curvature with a multiple of the Planck curvature\footnote{As discussed in  \cite{BodendorferOnTheCanonical}, leaving the homogeneous and isotropic sector, the two theories show different higher curvature corrections. However, the transition behaviour in the high curvature regime of the solution given in \cite{ChamseddineResolvingCosmologicalSingularities} qualitatively agrees with the numerical analysis of \cite{AshtekarLoopQuantumCosmology}.}. 

Now, as for the 5d Planck scale quantum corrected metric considered in Eq. (\ref{qmetric2}), the bulk metric defined in Eqs. (\ref{qmetric3a}, \ref{qmetric3b}) is not singular at $t=0$. Hence, as argued in Sec. \ref{5dscale}, following the usual AdS/CFT logic we take a 5d bulk quantum gravity point of view and replace $\lambda$ with a 5d effective scale $z\lambda$ in (\ref{CM}), thus leading us to the metric (\ref{qmetric3a}, \ref{qmetric3b}).

The metric (\ref{qmetric3b}) allows to include Kasner transitions in our analysis. Indeed, for $\left|\frac{t}{z\lambda}\right|\gg1$, the following approximation holds

\be\label{asymp}
\sinh^{-1}\left(\frac{t}{z\lambda}\right)\simeq\pm\log{\left| 2 \frac{t}{z\lambda}\right|}\;,
\ee
where the plus and minus signs are to be taken for $t\gg z\lambda$ and $t\ll -z\lambda$, respectively. The scale factor (\ref{qmetric3b}) then simplifies to

\begin{equation}\label{classlim}
a(t,z)^2\simeq\frac{a_{ext}^2}{\lambda^{2p}} \,\left(\frac{2}{\lambda z}\right)^{\pm 2 \Delta p}\, t^{2p_\pm}\;,
\end{equation}
with Kasner exponents

\begin{equation}\label{kasnerexp}
p_\pm=p\pm\Delta p\;.
\end{equation}

\begin{figure} [t!]
	\centering
	\includegraphics[width=8cm]{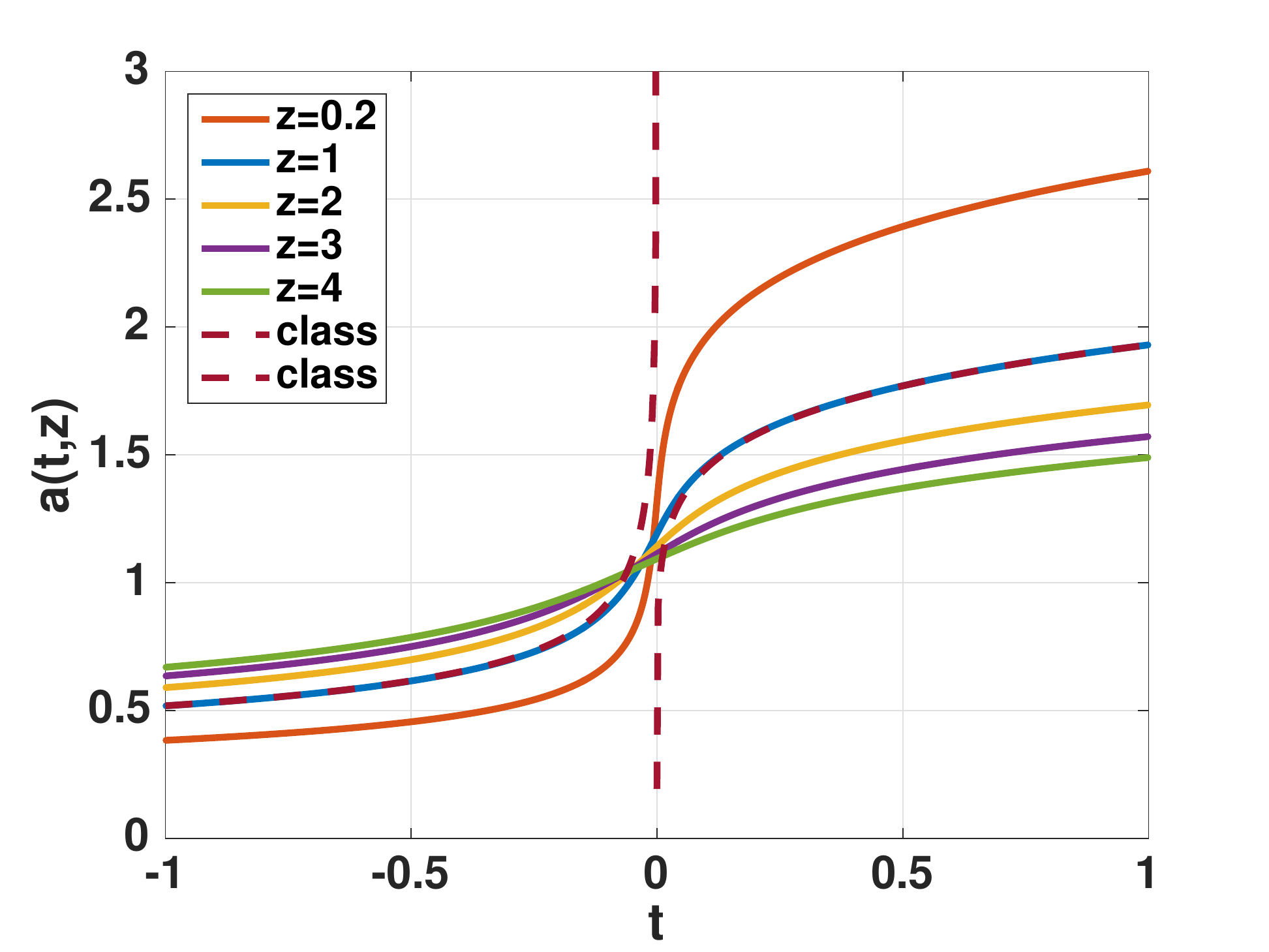}
	\caption{Plot of the scale factor \eqref{qmetric3b} vs. $t$ for $\lambda=0.06, p=-\frac{1}{16}, \Delta p=\frac{3}{16}$. Away from the immediate vicinity of the bounce point ($t=0$) the different lines, which correspond to different values of $z$, agree with the classical behaviour (explicitly reported for $z=1$ in dashed lines), and show a smooth transition during the bounce.}
	\label{fig:scalefactor}
\end{figure}
\noindent
Therefore, as also shown in Fig. \ref{fig:scalefactor}, asymptotically far in the past and in the future, the scale factor (\ref{qmetric3b}) for fixed non-zero $z$ reduces to that of the classical 4d-Kasner metric (up to the scaling factor $(\frac{2}{\lambda z})^{\pm2\Delta p}$ to be discussed below) with Kasner exponents given in (\ref{kasnerexp}). Quantum effects become dominant for $\left|\frac{t}{z\lambda}\right|\ll1$. In such a regime, we have

\be
a(t,z)^2\simeq a_{ext}^2z^{2p}\left(1+2\Delta p\frac{t}{z\lambda}\right)\;,
\ee 
and the metric describes a regular bounce (around $t=0$) during which the exponents characterising the Kasner universe are changing from $p_-$ to $p_+=p_-+2\Delta p$ after the bounce. This qualitatively agrees with the behaviour expected from LQC predictions \cite{GuptQuantumGravitationalKasner}\footnote{It turns out that this is the only possible transition behaviour in accordance with the correct classical limit for both early and late times \cite{WilsonEwingTheloopquantum}.}. Obviously for $\Delta p=0$ there are no Kasner transitions (i.e., $p_-=p_+$), and coherently the metric (\ref{qmetric3a}, \ref{qmetric3b}) reduces to (\ref{qmetric2}).

Although, the metric (\ref{qmetric3a}, \ref{qmetric3b}), seems to be a good candidate for the 5d-quantum metric, it has the following problems.
First, it is not a solution of the 5d-Einstein equations up to quantum corrections of order $\mathcal{O}(\lambda)$. This a priori unexpected issue comes about as follows: 

As shown in Fig. \ref{fig:scalefactor}, a scale factor of the type \eqref{qmetric3b} connects two Kasner branches, one with a negative and one with a positive exponent. They are matched such that the solution is monotonically increasing (or decreasing). In order to achieve this, one needs to scale the prefactors of $t^{2p_+}$ as well as $t^{2p_-}$ relatively to each other. In the classical limit, which is achieved either by $\lambda \rightarrow 0$ or here equivalently for the 4d-part by $z \rightarrow 0$, the relative scaling diverges as shown in Eq. \eqref{classlim}. At finite $z$, this effect is responsible for \eqref{qmetric3a} not solving the classical Einstein equations at large $t$, since this matching during the bounce induces a $z$-dependence in $a(t,z)$ which persists large $t$.

This indicates the second problem: Since Eq. \eqref{classlim} diverges for $\lambda=0$, it is not possible to recover the 4d classical Kasner geometry globally. Moreover, since Eq. \eqref{classlim} diverges also for $z=0$, the boundary limit is not well-defined. One may try to circumvent this issue by modifying the metric \eqref{qmetric3b}, e.g., as
\begin{figure} [t!]
	\centering
	\includegraphics[width=8cm]{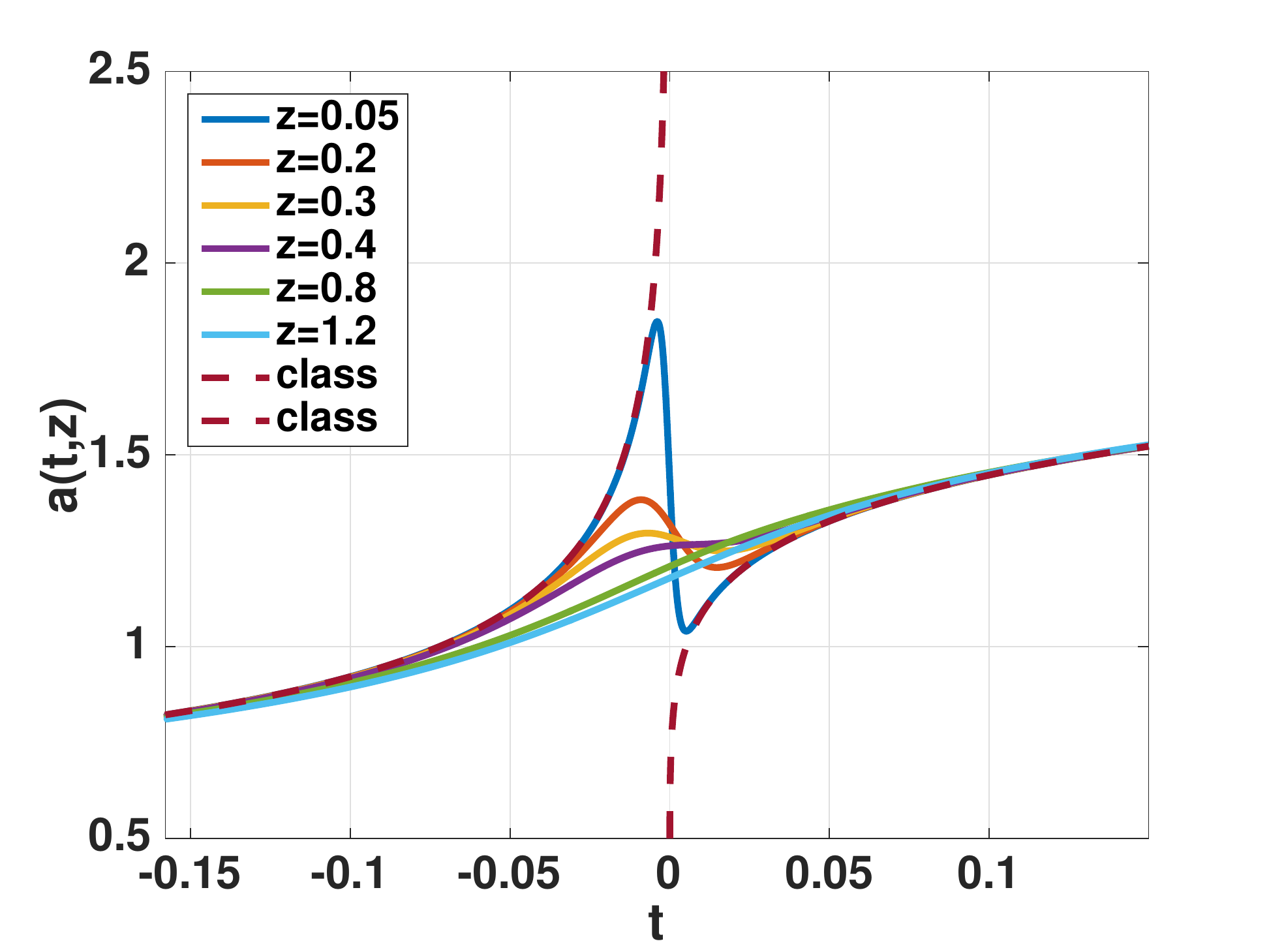}
	\caption{Plot of the scale factor \eqref{qmetric4} vs. $t$ for $\lambda=0.06, p=-\frac{1}{16}, \Delta p=\frac{3}{16}$. Away from the immediate vicinity of the bounce point ($t=0$) the different lines, which correspond to different values of $z$, agree with the classical behaviour (dashed lines) for all values of $z$, and show a smooth transition during the bounce. Nevertheless, this deviates qualitatively from the LQC behaviour.}
	\label{fig:scalefactorz}
\end{figure}
\be\label{qmetric4}
a(t,z)^2=\frac{a_{ext}^2}{\lambda^{2p}}\left(t^2+\lambda^2 z^2\right)^p\exp{\left[2\Delta p\sinh^{-1}\left(\frac{t}{z\lambda}\right)\right]}(\lambda z)^{2\Delta p\,\tanh{\left(\frac{t}{z\lambda}\right)}}\;,
\ee

\noindent
where the factor $(\lambda z)^{2\Delta p\tanh{\left(\frac{t}{z\lambda}\right)}}$ allows to recover the proper boundary (classical) limit. Indeed for $\left|\frac{t}{z\lambda}\right|\gg1$, taking into account that $\tanh{\left(\frac{t}{z\lambda}\right)}\simeq\text{sign}(t)$ ($z,\lambda\geq0$) together with Eq. \eqref{asymp}, the scale factor \eqref{qmetric4} reduces to
\be
a(t,z)^2\simeq a(t)^2=\frac{a_{ext}^2}{\lambda^{2p}}\,t^{2p_\pm}\;,
\ee

\noindent
which does not depend on $z$, and the classical Kasner case is recovered in the double scaling limit $\lambda\rightarrow0, a_{ext}/\lambda^p\rightarrow1$.
This modification ensures the correct classical and boundary limits. Fig. \ref{fig:scalefactorz} shows the time dependence of the scale factor for different values of $z$. In contrast to the previous case for all values of $z$ the same classical solution is approached for large $t$. However, due to the lack of relative rescaling of the prefactors, we necessarily have a non-monotonous behaviour around $t=0$ for small $z$. Indeed, around $t=0$, the scale factor can be written as

\be
a(t,z)^2\simeq a^2_{ext}\,z^{2p}\left[1+2\Delta p\frac{t}{\lambda z}\bigl(1+\log{(\lambda z)}\bigr)\right]\;,
\ee

\noindent
and, as reported in Fig. \ref{fig:scalefactorz}, for small $z$ the logarithmic term is responsible for the change in sign of the slope around $t=0$. This differs qualitatively from the LQC behaviour.

Unlike \eqref{qmetric3b}, the metric \eqref{qmetric4} is a solution of the 5d-Einstein equations in zeroth order in $\lambda$. Indeed, as already remarked in Sec. \ref{sec:metric1}, in order to have a plausible embedding of the 4d metric into the 5d bulk space, two conditions have to be satisfied: First, the boundary limit is well defined and solves the 4d-Einstein equations; second, the $z$-derivative of the scale factor vanishes in zeroth order in $\lambda$. On the one hand, for the metric \eqref{qmetric3b} the proper boundary limit is not recovered and the $z$-derivative does not vanish in zeroth order in $\lambda$ (see Eq. \eqref{zderivative2} below). On the other hand both conditions are satisfied for the metric \eqref{qmetric4}. 
However, the Kretschmann scalar for $t=0$ is not constant in $z$ and furthermore diverges for $z\rightarrow 0$ due to a blow-up of the $\mathcal O(\lambda)$ coefficient, as can be checked using computer algebra. Therefore, unlike \eqref{qmetric3b}, we cannot interpret \eqref{qmetric4} as an effective metric of a quantum gravity theory that resolves singularities by means of a limiting 5d bulk curvature.

Hence, finding a plausible metric that cures all problems at once turns out to be a highly non-trivial task. In principle the effective form of the quantum corrected metric should be derived from full quantum gravity, which would amount to set up 5d quantum Einstein equations and to extract an effective metric, at least in a suitable midisuperspace. We will leave this for further research. Here in this work, we decided to continue, as a case study, with the metric \eqref{qmetric3b}\footnote{We performed our numerical analysis also for the metric \eqref{qmetric4}. Without entering into details, the main conclusions are not affected by the modification in Eq. \eqref{qmetric4} and the finite-distance pole of the two-point correlator is resolved also for this metric.} since, unlike \eqref{qmetric4}, it shows qualitatively the behaviour we expect from LQC in the 4d part of metric, the Kretschmann scalar exhibits the right behaviour, and it is also symmetric under the scaling symmetry \eqref{scaling}. As we are interested here in studying the two-point correlator, we use (\ref{qmetric3b}) only as an example to check the genericity of the absence of the finite-distance pole. We do not claim that it is the correct effective metric. 

\subsubsection{Solution of the Geodesic Equations}

The same procedure as for the metric \eqref{qmetric2} is now repeated for the case of Kasner transitions, i.e., for the metric defined in Eqs. \eqref{qmetric3a} and \eqref{qmetric3b}. The derivatives of $a(t,z)^2$ entering Eqs.\eqref{eq:geo-ivp} are now given by:
\begin{align}
&\partdif{(a^2)}{t} = 2 a(t,z)^2 \left[ \frac{p t}{t^2 + \lambda^2 z^2} + \Delta p  \frac{1}{\sqrt{t^2+\lambda^2 z^2}} \right] \;, 
\\
&\partdif{(a^2)}{z} = 2 a(t,z)^2 \left[ \frac{p \lambda^2 z}{t^2 + \lambda^2 z^2} - \frac{\Delta p}{z} \frac{t}{\sqrt{t^2+\lambda^2 z^2}}  \right] \;. 
\label{zderivative2}
\end{align}

\begin{figure} [t!]
	\centering
	\includegraphics[width=9.5cm]{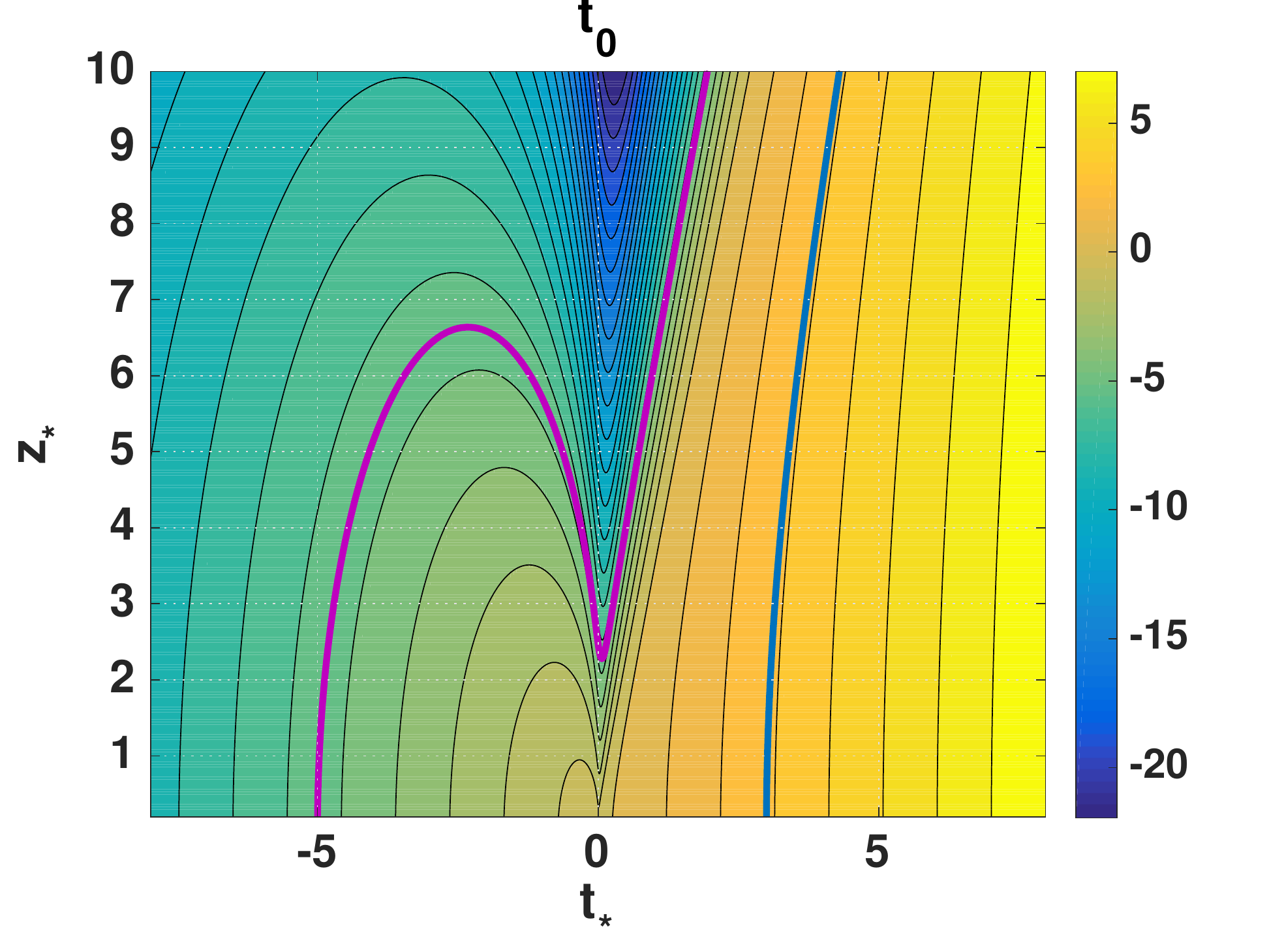}
	\caption{Colour plot of $t_0(t_*, z_*)$ for the metric \eqref{qmetric3a} with $t_0 = -5$ (purple) and $t_0=3$ (blue), $\lambda = 0.06$, $p=-1/16$, $\Delta p = 3/16$ , $a_{ext} = \lambda^p$. Passing from negative to positive $t$, there is a transition from $p_- = -1/4$ to $p_+ = 1/8$. We see two kinds of solutions: The ones starting at negative $t_0$, which are bent towards the (resolved) singularity and eventually passing it (purple) and the ones starting at positive $t_0$, which are bent away from the (resolved) singularity (blue).}
	
	\label{fig:t0contour2}
\end{figure}

\begin{figure} [t!]
	\centering
	\includegraphics[width=9.5cm]{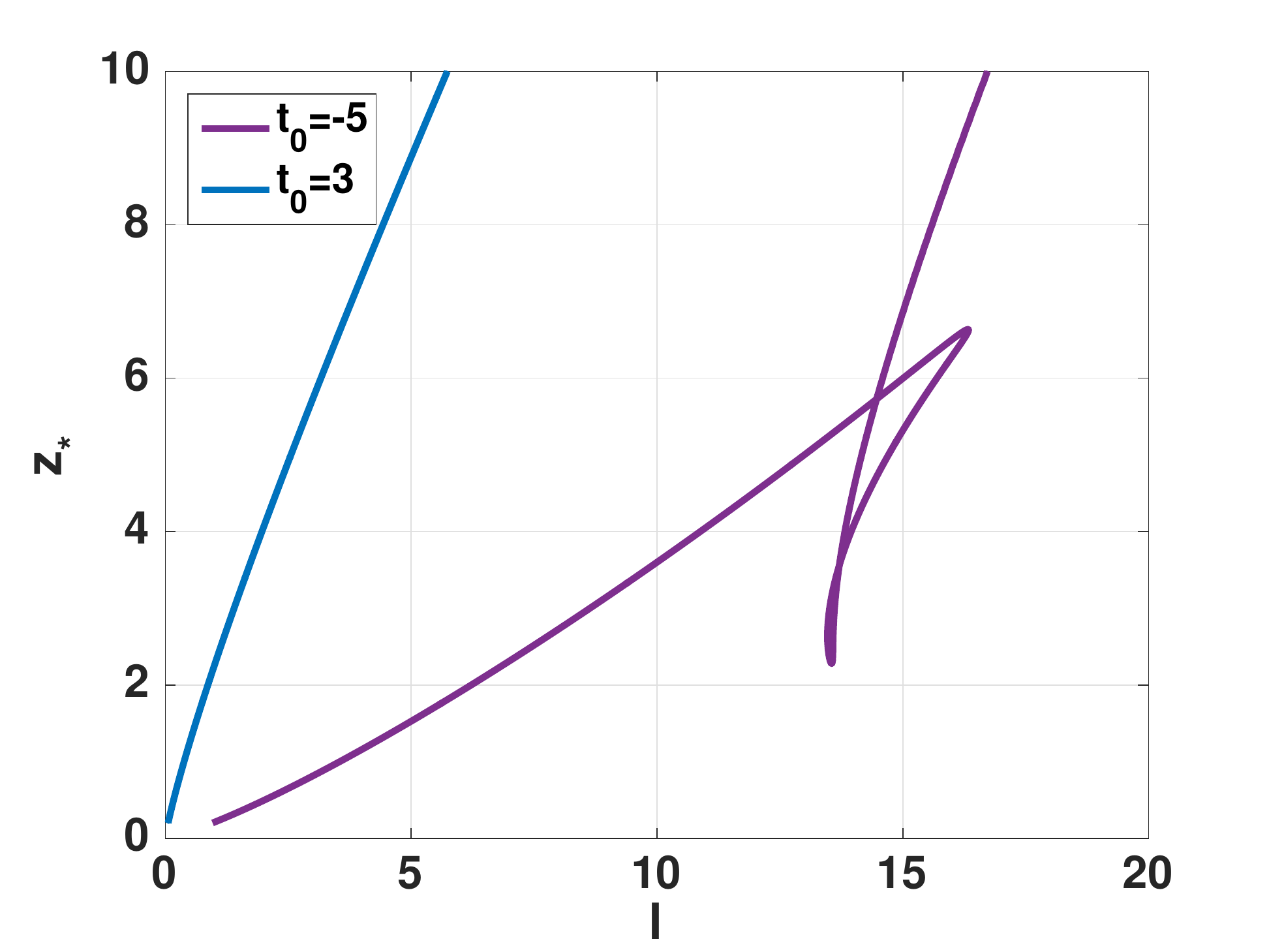}
	\caption{$z_*$ vs. $l$ for $t_0 = -5$ (purple) and $3$ (blue) and $\lambda = 0.06$, $p=-1/16$, $\Delta p = 3/16$ , $a_{ext} = \lambda^p$.}
	
	\label{fig:zstarvsl2}
\end{figure}

\noindent
The corresponding colour plot of $t_0$ is shown in Fig. \ref{fig:t0contour2}.
Unlike the previous case, the range of $t_*$ includes also negative values so that we have now the possibility to study geodesics passing the resolved singularity at $t = 0$. In Fig. \ref{fig:t0contour2} the level lines for $t_0=-5$ (purple) and $t_0 = 3$ (blue) are singled out. Again, there is a turning point in $z_*$, close to $t_* = 0$. Moreover, there are solutions that start at a negative $t_0$, pass through the resolved singularity, have their turning point at positive $t_*$ and come back to the $t_0$-value they started from. For example, there are points of the level line for $t_0 = -5$ that reach positive $t_*$. Furthermore, it is possible to observe the dominant behaviour of the different values of $p$. In the region $t_* < 0$, the plot looks similar to Fig. \ref{fig:contour1} (a), just mirrored. For $t > 0$, the plot has exactly the behaviour of positive $p$ solutions. As expected, the transition area is smoothened out. Coherently with the analysis of \cite{EngelhardtFurtherHolographicInvestigations}, geodesics with $t_0 < 0$ are bent towards $t=0$, while geodesics with $t_0>0$ are bent away from $t=0$. Accordingly, as reported in Fig. \ref{fig:zstarvsl2}, the former correspond to the behaviour of $z_*(l)$ described by the purple line, while the latter to the blue line.
In both cases, $z_*$ never hits $z_*=0$ for finite values of $l$. However, since we are interested in probing the resolved bulk singularity, in what follows we shall focus on solutions of the first kind ($t_0 < 0$).

\subsubsection{Renormalised Length}

\begin{figure} [t!]
	\centering
	\includegraphics[width=10cm]{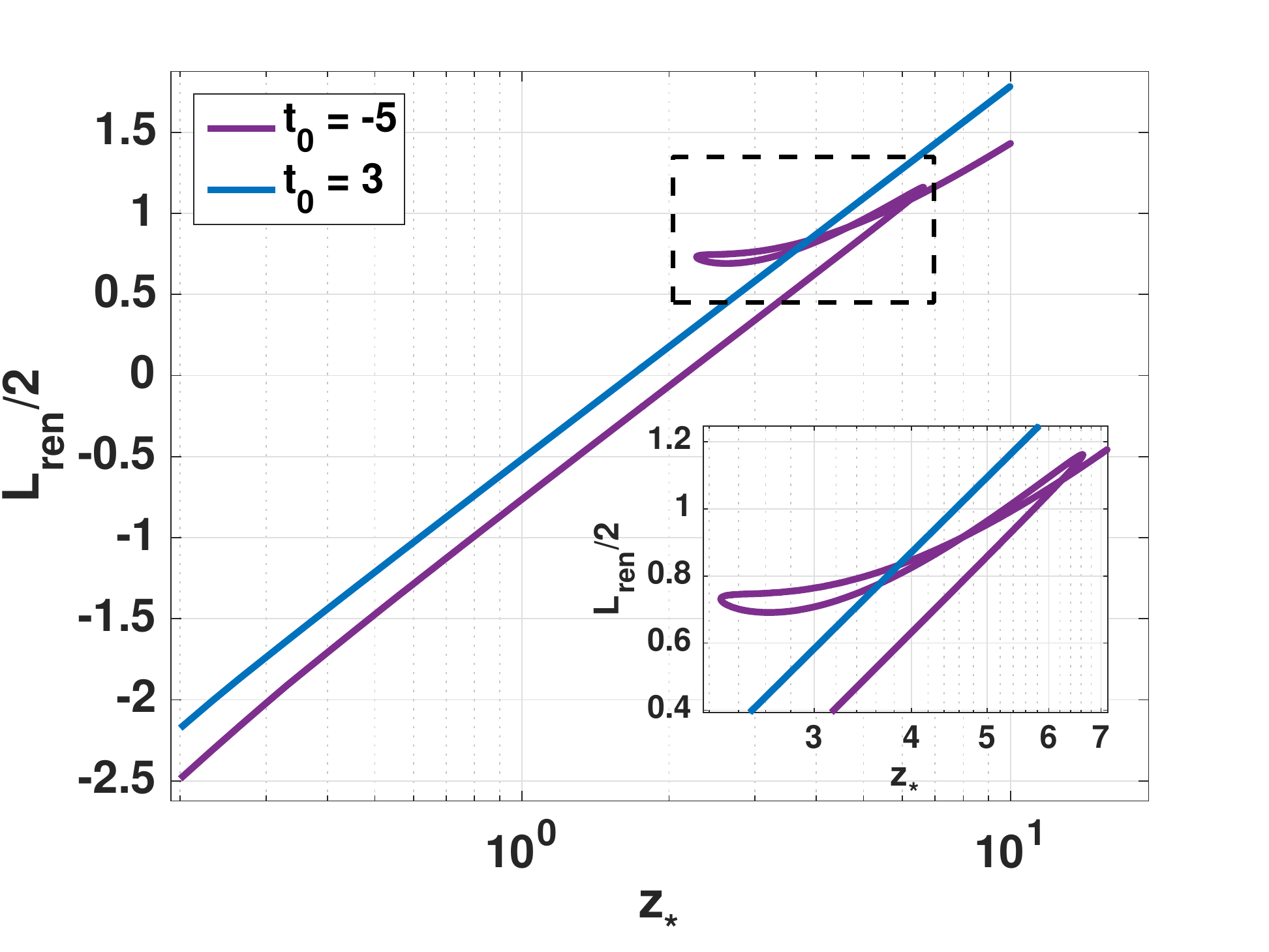}
	\caption{Log-scale plot of $L_{\text{ren}}$ vs. $z_*$ for the metric \eqref{qmetric3a} for $t_0 = -5$ (purple) and $t_0=3$ (blue), $\lambda = 0.06$, $p=-1/16$, $\Delta p = 3/16$ , $a_{ext} = \lambda^p$. Again an asymptotic $\log$-behaviour is visible.}
	
	\label{fig:Lrenvszstartrans1}
\end{figure}

Similarly to Sec. \ref{sec:metric1}, Fig. \ref{fig:Lrenvszstartrans1} shows a plot of the renormalised length $L_{\text{ren}}$ against $z_*$ in a logarithmic scale.
Up to the above-mentioned error occurring in the lower region of the line, we again have linear asymptotic behaviour, i.e., $L_{\text{ren}} \propto \log(z_*)$, staying away form $z_*=0$ for finite $l$.
For such kind of solutions (purple line in Figs. \ref{fig:zstarvsl2}, \ref{fig:Lrenvszstartrans1}) $L_{\text{ren}}$ remains finite for $l \neq 0$, and hence the two-point correlator does not diverge. For completeness, we also include in Fig. \ref{fig:Lrenvszstartrans1} the plot of a positive $t_0$ solution (blue). This kind of solutions are bent away from the region where quantum effects dominate, hence they behave as in the classical case and there is no pole in the correlator.

\section{Conclusions and Outlook}\label{conclusion}

In this paper, we have studied the effect of resolved cosmological singularities in dual field theories via the AdS/CFT correspondence using numerical techniques. Specifically, we have extended previous results \cite{BodendorferHolographicSignaturesOf} beyond a certain simplifying assumption that made analytic computations possible. 
As argued, it was non-trivial that the qualitative form of the results remained the same.

Our results show that the simplest possible ideas for resolving cosmological singularities lead to sensible results in the dual field theory that could in principle be checked by an independent calculation. Doing so would be of great interest as this would allow to study the question of whether specific proposals for non-perturbative canonical quantum gravity could be seen as a certain subsector of string theory non-perturbatively defined via its dual field theory. 

Due to the BKL conjecture \cite{BelinskiiOscillatoryApproachTo}, stating that the spacetime dynamics near generic spacelike singularities decouples into homogeneous patches, the relevance of our computation may well go beyond the specific Kasner-AdS solution considered here.

In future work, it would be interesting to extend the present calculation to more general spacetimes, e.g. by substituting the Kasner spacetime in $ds^2_5$ by some other 4d spacetime. Of particular interest are spacetimes containing black holes, on which recent efforts in loop quantum gravity have been focussed \cite{HaggardBlackHoleFireworks, CorichiLoopQuantizationOf, DeLorenzoImprovedBlackHole, OlmedoFromBlackHoles}. Clearly, it must be the aim to find scenarios where we can independently understand the dual field theory, e.g. via lattice gauge theory.

\section*{Acknowledgements}
The authors were supported by an International Junior Research Group grant of the Elite Network of Bavaria. Discussions with Andreas Sch\"afer and John Schliemann are gratefully acknowledged.

%\bibliographystyle{utphysmendeley}
%\bibliography{library}

\end{document}